\documentclass[amsmath,amssymb,aps,twocolumn,floatfix]{revtex4-2}
\usepackage{adjustbox}
\usepackage{float}
\usepackage{physics}
\usepackage{subfigure}
\usepackage{graphicx}
\usepackage{dcolumn}
\usepackage{bm}
\usepackage{hyperref}
\usepackage[usenames]{color}
\usepackage{tikz}
\usepackage{amssymb}
\usepackage{amsmath}
\usepackage{multirow}
\usepackage{array}
\usepackage{booktabs}
\usepackage{threeparttable}
\usepackage{siunitx}
\usepackage{multirow}

\begin{document}
\newcommand{\NMcomment}[1]{{\bf \textcolor{red}{ #1}}}
\newcommand{\ALcomment}[1]{{\bf \textcolor{blue}{ #1}}}
\newcommand{\lb}{\left [}
\newcommand{\rb}{\right ]}
\newcommand{\lpipe}{\left |}
\newcommand{\rpipe}{\right |}
\newcommand{\pd}[2]{\frac{\partial #1}{\partial#2}}
\newcommand{\der}[2]{\frac{d #1}{d #2}}
\newcommand{\secpd}[2]{\frac{\partial^2 #1}{\partial #2^2}}
\newcommand{\secder}[2]{\frac{d^2 #1}{d #2^2}}

\sisetup{tight-spacing=true,text-series-to-math,propagate-math-font}
\preprint{APS/123-QED}

\title{Homonuclear ultracold elastic $s$-wave collisions of alkali atoms via multichannel quantum defect theory }

\author{Alyson Laskowski}
 \email{Current address: University of Illinois Chicago, alasko5@uic.edu}
\author{Nirav P. Mehta}%
 \email{nmehta@trinity.edu}
\affiliation{%
 Department of Physics and Astronomy, Trinity University, San Antonio, Texas 78212, USA\\
}%

\date{\today}

\begin{abstract}
  Multichannel quantum defect theory (MQDT) provides a powerful toolkit for describing and understanding collisions of cold alkali atoms. Various MQDT approximations differ primarily in how they characterize the so-called short-ranged $K$-matrix, ${\mathbf K}_{\text{sr}}$, which encapsulates the short-ranged, high-energy physics into a handful of low-energy parameters that exhibit simple and smooth dependence on energy and field. Here, we compare three different methods for computing ${\mathbf K}_{\text{sr}}$ for homonuclear collisions of alkali atoms, from lithium to cesium. The MQDT calculations are benchmarked against numerically converged coupled-channels calculations that use a log-derivative propagator out to the asymptotic region. We study how well these approximations reproduce positions of $s$-wave magnetic Feshbach resonances, comparing to experiment where possible, and identify the limitations of various approximations.
\end{abstract}

\maketitle

\section{Introduction}
\label{sec:intro}
The ability to control the scattering length by tuning an applied magnetic field
in the vicinity of a magnetic Feshbach resonance has now become a standard tool
in experimental ultracold physics~\cite{chin_feshbach_2010}. For example,
manipulation of the scattering length in this manner plays a key role in the
realization of strongly interacting many-body systems~\cite{bloch2008many}. It
enables the creation of loosely bound molecules via Feshbach association, which
is the first step in the formation of deeply bound molecules by subsequent
stimulated Raman adiabatic passage~\cite{regal2003creation}.  Control of the
two-body scattering length in this manner has played a key role in the study of
Efimov physics~\cite{berninger2011universality, kraemer2006evidence,ferlaino2009evidence,huang2014observation,roy2013test,dyke2013finite}. Theoretical
developments have kept pace with experiment in predicting and understanding the
properties of magnetic Feshbach resonances~\cite{chin_feshbach_2010}, and one of
the most powerful theoretical tools that has been brought to bear upon the
problem is multichannel quantum defect theory (MQDT).

MQDT provides a powerful
formalism for computing and understanding the field and energy
dependence of collisional cross sections in ultracold systems.
It has a long history, with seminal contributions made by many
authors~\cite{seaton1958quantum,seaton1966quantumI,seaton1966quantumII,
  greene1979general,greene1982general,mies1984multichannelI,mies1984multichannelII,
  gao1998solutions,gao1998quantum,gao2000zero,gao2001angular,
  gao2005multichannel}. Various formulations differ
significantly in notation and scope, but not in spirit. MQDT at its heart leverages a
separation of energy and length scales in order to simplify the
calculation of low-energy observables.  It is in this sense an
``effective theory'' similar in spirit to modern renormalization techniques and
effective field theories. In its application to ultracold atomic
collisions, it can be made to agree with coupled channels
calculations to a numerical accuracy approaching exactness.

The strength and nature of ultracold collisions depends on the separation distance $R$ between atoms. At small $R$, the wells of the ground state spin-singlet and spin-triplet Born-Oppenheimer potentials are many orders of magnitude deeper than any other relevant energy scale, including those of the long-range van der Waals tail and one-atom hyperfine-Zeeman interactions. This robust separation of energy and length scales enables one to treat the collision in stages. First, one solves the
short-range problem to determine a short-range $K$-matrix, ${\mathbf K}_{\text{sr}}$, which is
defined with respect to energy-analytic reference functions $\{\hat{f},\hat{g}\}$ that are solutions to the long-range (e.g. van der Waals) potential
common to all collision channels. Then one treats the long-range
physics using the methods of MQDT, which involve accounting for (1) the phase
accumulated in the long-range potential by $\{\hat{f},\hat{g}\}$---both with respect to each other and with respect to a pair of
energy-normalized solutions $\{f,g\}$, (2) the energy-normalization of
$\{f,g\}$, particularly when expressed in terms of the energy-analytic
pair $\{\hat{f},\hat{g}\}$, and (3) the reflected amplitude from closed
channels. 

The short-range $K$-matrix is viewed as “input” into the machinery of MQDT, encoding information about the short-range physics relevant to low-energy (near threshold) observables. Moreover, ${\mathbf K}_{\text{sr}}$ exhibits a smooth and simple dependence on both energy and magnetic field. Therefore, it only needs to be calculated on a coarse grid of energy and field values to provide a complete description of the short-range physics. The frame transformation (FT)~\cite{burke1998multichannel} provides a powerful
tool for approximating ${\mathbf K}_{\text{sr}}$ by writing it in terms of the singlet 
and triplet quantum defects $\mu_S$ and a sum over the spin singlet ($S=0$) and 
spin triplet ($S=1$) projection operators.  A re-coupling then rotates
${\mathbf K}_{\text{sr}}$ into the field-dressed hyperfine basis that diagonalizes the long-range
Hamiltonian.  In the limit that the hyperfine and Zeeman splittings vanish, the
frame transformation becomes essentially exact, limited only by the quality of the energy-analytic reference functions. We consider two variations of the frame
transformation: (1) the energy independent frame transformation (EIFT), which
requires only the zero-energy quantum defects to compute ${\mathbf K}_{\text{sr}}$, and (2) the
energy-dependent frame transformation (EDFT), which requires the quantum defects
on a course grid of energy spanning the separation of two-body collision
thresholds determined by the hyperfine-Zeeman energies.

A number of studies~\cite{hanna2009prediction,gao2011analytic,cui2018broad} have
utilized an energy independent frame transformation to build essentially a
three-parameter MQDT that requires only the singlet and triplet scattering
lengths $a_S$ and $a_T$, and the leading dispersion coefficient $C_6$. In such a
scheme, $a_S$, $a_T$ and $C_6$ may be considered tunable parameters that can be
adjusted to reproduce low-energy observables such as the positions of certain
Feshbach resonances. The simplicity of this approach gives it enormous
predictive power, as demonstrated by a recent study that identified a very large
number of ``broad'' Feshbach resonances~\cite{cui2018broad}. The present study
places such frame transformation calculations in context by providing direct
comparisons to more accurate implementations of MQDT, and also to numerically converged
coupled channels calculations, which we take here to be ``exact".

This paper is structured as follows. In Section~\ref{sec:collisionmodel}, we discuss our model of alkali collisions, including the interaction Hamiltonian and field-dressed 
hyperfine-Zeeman basis. We describe the various Born-Oppenheimer potentials adapted for this work, discussing their properties and any necessary modifications made for the present calculations. Section~\ref{sec:mqdt-theory} provides a brief overview of MQDT for ultracold collisions along with explanations of EIFT and EDFT. Our results, including the positions of $s$-wave resonance and zero crossings for particular collision channels of each species, are presented in Section~\ref{results}. 

We show that when one obtains ${\mathbf K}_{\text{sr}}$ from a rigorous
boundary condition on a multichannel short-ranged solution---what we shall refer to as the ``MQDT" calculation, the low-energy
scattering observables agree, nearly exactly, with converged coupled-channels (CC)
calculations using Johnson's log-derivative
propagator~\cite{johnson1973multichannel}. The agreement between MQDT and CC
calculations, however, is only possible if the model potential energy functions
for the singlet and triplet configurations reliably converge to the long-range
dispersion form Eq.~(\ref{eq:vlr}) at separation distances where all collision
channels are locally open. We also find that frame transformation approximations for ${\mathbf K}_{\text{sr}}$ provide an excellent
description of lighter alkali species, especially lithium, but become
progressively unreliable for heavier species in which the hyperfine-Zeeman
splitting is much larger, and the energy-dependence of the quantum defects over
the necessary range of energy is appreciable.

Finally, it is worth mentioning that while analytical solutions to the Schrodinger equation for potentials that vary as $R^{-6}$ have been formulated by Gao~\cite{gao1998solutions}, we
opt instead to use of the numerical approach proposed
in~\cite{yoo1986implementation}, namely the Milne phase amplitude
method~\cite{milne1930numerical}, to compute the energy-analytic reference
functions that play a key role in MQDT. This approach is, for our purpose,
simpler and more versatile since it is applicable to the case of a more general
long-range potential that includes higher order dispersion terms---including these long-range dispersion terms reduces the energy dependence of the quantum defects and generally improves the MQDT. It is also, in our modest view, simpler to implement than the rather complicated analytical solution of~\cite{gao1998solutions}.

To a new student of MQDT, the literature can be daunting. In the process of this
work, we have relied heavily on Refs.~\cite{burke1998multichannel,ruzic2013quantum} to gain an understanding of MQDT methods, particularly as they relate to ultracold atomic collisions. The appendix of Ref.~\cite{mcalexander2000collisional} provides useful expressions for matrix elements relevant to the hyperfine-Zeeman hamiltonian, and Ref.~\cite{yoo1986implementation} provides a good starting point for computing the energy-analytic reference functions. 

\section{Model of Alkali Collisions}
\label{sec:collisionmodel}
Our model for ultracold collisions of alkali atoms follows closely that of
Ref.~\cite{stoof_spin-exchange_1988}. For two-body atomic scattering, one
generally writes the wavefunction as
$\Psi(R,\Omega)=R^{-1}\sum_i{\psi_i(R)\Phi_i(\Omega)}$, where $R$ is the
nuclear separation of the atoms and $\Omega$ is a collective coordinate describing all
angular and internal degrees of freedom.  The problem is reduced to a coupled
channels equation of the form
\begin{equation}
  \label{mcse}
  \sum_j{\left[\frac{\hbar^2}{2\mu}\left(-\frac{d^2}{dR^2}+\frac{\ell_j(\ell_j+1)}{R^2}\right)\delta_{ij}+V_{ij}\right]\psi_j}=E\psi_i.
\end{equation}
Here, $\mu$ is the reduced mass of the homonuclear dimer. The interaction matrix
${\bf V}(R)$ for two ultracold alkali atoms in a magnetic field is of the form
\begin{equation}
  \label{eq:Vfull}
  {\bf V}(R) = {\bf P}_0 V_0(R) + {\bf P}_1 V_1(R)  + \sum_{n=1}^2{\bf H}_n^{\text{HZ}},
\end{equation}
where ${\bf P}_0$ and ${\bf P}_1$ are the singlet and triplet projection
operators, and $V_S(R)$ are the Born Oppenheimer potentials
corresponding to the singlet ($S=0$) and triplet ($S=1$) molecular ground states $X^1
\Sigma_{g}^{+}$ and $a^3 \Sigma_{u}^{+}$, respectively. The matrix operator
${\bf H}_n^{\text{HZ}}$ is the combined hyperfine and Zeeman interaction for
each atom,
\begin{equation}
  \label{eq:HHZ}
  {\bf H}_n^{\text{HZ}} = {\left[\frac{A_n}{\hbar^2}
      \vec{\bf s}_n \cdot \vec{\bf i}_n +
      \frac{\mu_B}{\hbar}\left(g_{s} \vec{\bf s}_n + g_{ni} \vec{\bf i}_n\right) \cdot \vec{B} \right]},
\end{equation}
where $\vec{\bf i}_n$ and $\vec{\bf s}_n$ are the nuclear and electronic
spins of atom $n$, $A_n$ is the hyperfine coupling in the electronic
ground state, and $g_s$ and $g_{ni}$ are electron and nuclear $g$-factors
in units of the bohr magneton $\mu_B$.  We adhere to the convention
of Ref.~\cite{arimondo1977experimental} and define the $g$-factors to be of
the opposite sign as their corresponding magnetic dipole moments.
For convenience and clarity, a collection of the relevant parameters from
Ref.~\cite{arimondo1977experimental} is given in Table~\ref{tbl:couplings}.  
\begin{table}[t]
  \caption{\label{tbl:couplings} Hyperfine couplings and nuclear $g$-factors used in this work. Nuclear g-factors $g_i$ should be multiplied by the bohr magneton.}
  \begin{threeparttable}
    \begin{ruledtabular}
        \begin{tabular}{cccc}
          Atom             & $i$   & $g_i$              & $A_{\text{hf}}/h [\text{MHz}]$ \\
          \midrule
          $^6\text{Li}$    & $1$   & $-0.0004476540(3)$   & $152.1368407(20)$   \\
          $^7\text{Li}$    & $3/2$ & $-0.001182213(6)$    & $401.7520433(5)$    \\
          $^{23}\text{Na}$ & $3/2$ & $-0.0008046108(8)$   & $885.8130644(5)$    \\
          $^{39}\text{K}$  & $3/2$ & $-0.00014193489(12)$ & $230.8598601(3)$    \\
          $^{40}\text{K}$  & $4$   & $+0.000176490(34)$   & $-285.7308(24)$     \\
          $^{85}\text{Rb}$ & $5/2$ & $-0.0002936400(6)$   & $1011.910813(2)$    \\
          $^{87}\text{Rb}$ & $3/2$ & $ -0.0009951414(10)$ & $3417.34130642(15)$ \\
          $^{133}\text{Cs}$ & $7/2$ & $ -0.00039885395(52)$ & $2298.1579425$ \\
        \end{tabular}
    \end{ruledtabular}
  \end{threeparttable}
\end{table}
\begin{figure*}[t]
\includegraphics[width=6.5in]{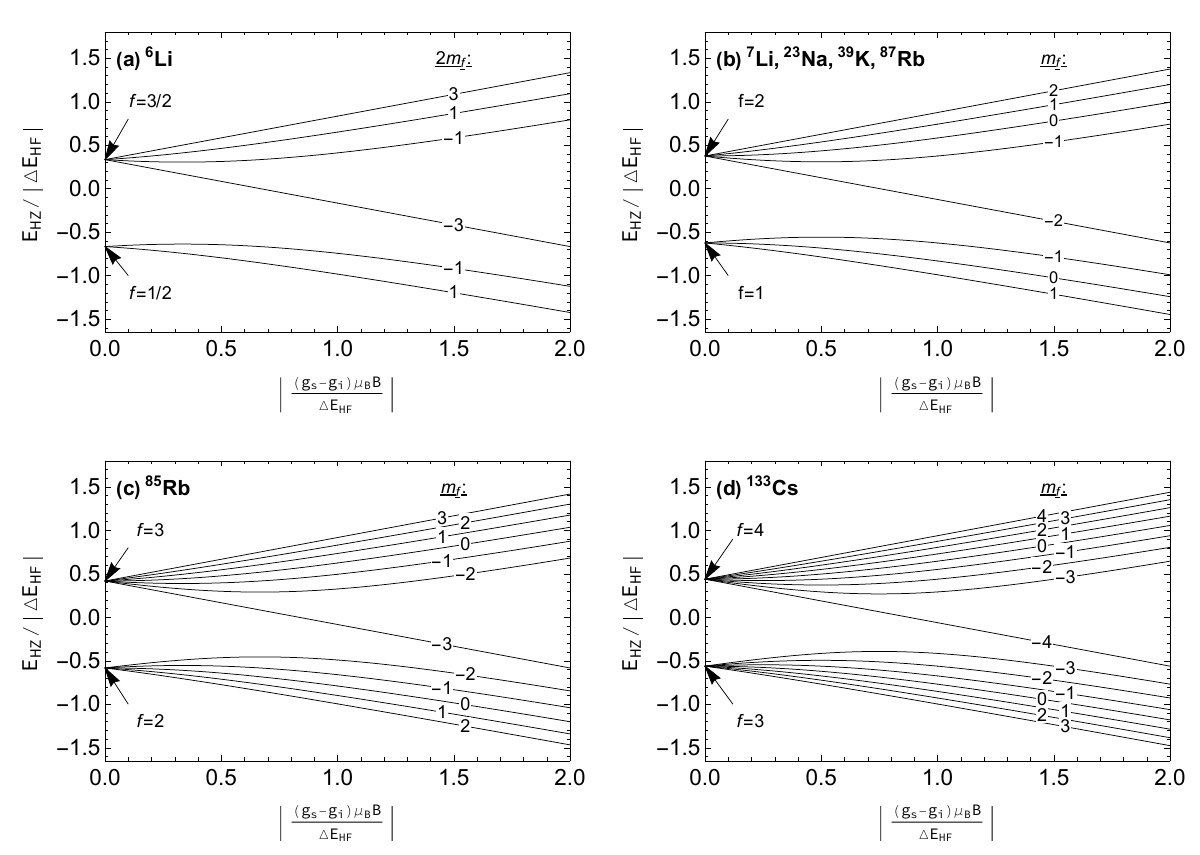}
\caption{\label{fig:BR} The one-atom Breit-Rabi energy spectrum for
  atomic species considered in this work.  The $f$ quantum number is
  only good in the zero-field limit.  At any field, $m_f$
  remains a good quantum number, but different $f$ levels are coupled. The curves are labeled by their $m_f$ quantum number, or in cases where $m_f$ is fractional, by $2m_f$.}
\end{figure*}

\begin{figure}[t]
\includegraphics[width=3.4in]{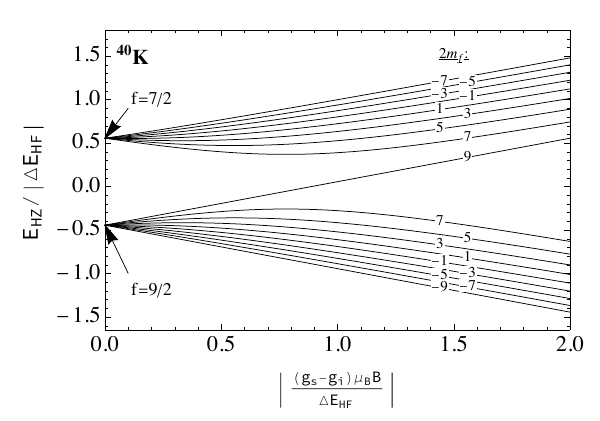}
\caption{\label{fig:BR-K} The one-atom Breit-Rabi energy spectrum for
  $^{40}\text{K}$, for which the hyperfine coupling is negative,
  leading to an inverted Breit-Rabi diagram.}
\end{figure}

The two-atom collision thresholds in a magnetic field are determined by the
eigenstates of ${\bf V}(R)$ in the limit $R\rightarrow\infty$, where 
$V_S(R)$ vanishes. These states are constructed by appropriately
symmetrizing the eigenstates of Eq.~(\ref{eq:HHZ}).  The hyperfine
interaction couples the nuclear spin $\vec{\mathbf i}$ and electronic spin 
$\vec{\mathbf s}$ of each atom, and is diagonal in the total atomic spin 
$\vec{\mathbf f}=\vec{\mathbf i}+\vec{\mathbf s}$. However, the Zeeman interaction couples states of different $f$, so that only the projection $m_f$ remains a good quantum number at finite
field. While the states of ${\bf H}_n^{\text{HZ}}$ can be found
analytically by the Breit-Rabi formula~\cite{breit1931measurement}---for a
detailed derivation, see~\cite{makrides2014multichannel}---in practice, we
compute the matrix elements of the Hamiltonian in Eq.~(\ref{eq:HHZ}) in the
hyperfine basis $\ket{f,m_f}$ and diagonalize the resulting matrix numerically.
Figure~\ref{fig:BR}, shows the energy levels of a single atom in a magnetic
field for all of the alkali species considered here except for $^{40}\text{K}$,
which has a negative hyperfine coupling constant that results in the inverted
diagram shown in Fig.~\ref{fig:BR-K}. As we discuss below, these energies will determine the two-atom collision thresholds.

The short-ranged physics ($R \lesssim 30 a_0$) of Eq.~(\ref{eq:Vfull}) is dominated by the very deep singlet and triplet potentials, while the long-range physics ($R \gtrsim 30 a_0$)
is controlled by the comparatively shallow van der Waals tail and 
weak hyperfine-Zeeman structure of the atoms. For $R \gtrsim 30 a_0$, the
off-diagonal elements of $V_{ij}(R)$ in Eq.~(\ref{mcse}) vanish, and the diagonal
elements are determined by the dispersion coefficients $C_6, C_8$ and $C_{10}$
\begin{equation}
  V_{ii}(R)-E_i^{\text{th}} \rightarrow V_{\text{LR}}(R)
\end{equation}
where $E_i^{\text{th}}$ are the collision thresholds and the long-range potential common to all channels is of the form
\begin{equation}
  \label{eq:vlr}
V_{\text{LR}}(R) = -\frac{C_6}{R^6}-\frac{C_8}{R^8}-\frac{C_{10}}{R^{10}} \;\; \text{for} \;R \gtrsim 30 a_0.
\end{equation}
The natural unit of length $\beta$ associated with $V_{\text{LR}}$, and the corresponding natural unit of energy $E_\beta$ are fixed by the depth of $V_{\text{LR}}$ at a separation distance $\beta$:
\begin{equation}
    \label{eq:betadef}
    E_\beta = \frac{\hbar^2}{2\mu \beta ^2} = |V_{\text{LR}}(\beta )|
\end{equation}
This definition reduces to twice the usual van der Waals length when
$C_8=C_{10}=0$,  $\beta_6 = (2\mu C_6/\hbar^2)^{1/4} = 2R_{\text{vdW}}$, and renders the dispersion coefficients unitless when expressed in these units.

\subsection{Field-dressed hyperfine basis}
For the two-atom system, we follow
Ref.~\cite{stoof_spin-exchange_1988} and represent the symmetry
requirements for identical bosons or fermions by defining the basis
kets as  
\begin{equation}
    \ket{\{\alpha \beta\}} = \frac{\ket{\alpha, \beta}\pm(-1)^{\ell}\ket{\beta, \alpha}}{\sqrt{2(1+\delta_{\alpha,\beta})}}
\label{eq:sym-basis}.
\end{equation}
where the Greek letters refer to the internal states of the individual
atoms. For example, $\ket{\alpha,\beta} = \ket{f_1,m_1,f_2,m_2}$ represents
atom 1 in hyperfine state $\ket{\alpha}=\ket{f_1,m_1}$ and atom 2 in
state $\ket{\beta}=\ket{f_2,m_2}$, while the $+(-)$ sign is taken for bosons (fermions). We neglect in this work the magnetic dipole-dipole interaction, so the $s$-wave remains decoupled from higher partial waves. Furthermore, the total $M_F=m_{f_1}+m_{f_2}$ remains a good quantum number at finite field. Each calculation presented here is specified by a particular $M_F$, within which the lowest one-atom states can be read by the Breit-Rabi graphs.

The properly symmetrized eigenstates of the two-atom hyperfine-Zeeman
Hamiltonian comprise the ``field-dressed'' basis, constructed as a linear
combination of symmetrized atomic hyperfine states 
\begin{equation}
  \label{eq:fielddressing}
\ket{i}=\sum_{\{\alpha\beta\}}{C_{\{\alpha\beta\}}^{i}\ket{ \{\alpha\beta\} }}.
\end{equation}
The scattering thresholds correspond to the elements of the diagonal
matrix
\begin{equation}
  \label{eq:Ethresh}
  {\bf E}_{\text{th}}={\bf C}^{\text{T}}(B){\bf
    H}^{\text{HZ}}{\bf C}(B),
\end{equation}
 where ${\bf C}(B)$ is the field-dependent rotation comprised of the eigenvector
elements $C_{\{\alpha\beta\}}^{i}$. We express and solve Eq.~(\ref{mcse}) in the
field-dressed spin basis given by Eq.~(\ref{eq:fielddressing}).

The scattering cross section is determined by matching the solutions to
asymptotic Bessel functions in the limit $R\rightarrow \infty$. In our
calculations, because we neglect the weak, long-ranged magnetic dipole-dipole
interaction, we match at a radius $R$ much larger than the natural length $\beta$, where both the singlet $V_0(R)$
and triplet $V_1(R)$ potentials become negligible, and the two-atom interaction
is reduced to a sum of one-atom terms: $ \lim_{R\rightarrow \infty}{\bf V(R)} =
{\bf H}^{\text{HZ}}=\sum_{n=1}^2{{\bf H}^{\text{HZ}}_n}$. In practice, $R\approx 20 \beta$ is sufficiently large to
ensure that the van der Waals tail is negligible. We consider only $s$-wave
collisions in this work, but a larger matching radius may be necessary for
higher partial waves, particularly at threshold energies. 

\subsection{Singlet/Triplet Potentials}
\label{subsec:STPots}

A great deal of effort has been expended by many
authors~\cite{knockel2004high,salumbides2008improved,leroy2006accurate,leroy2007new,leroy_accurate_2009, dattani_dpf_2011,
  knoop_feshbach_2011, falke_potassium_2008, strauss_hyperfine_2010,
  baldwin2012improved,coxon2010ground,sovkov2017re} in the development of state-of-the-art Born-Oppenheimer potential curves
for alkali dimers in the spin singlet ($X^1\Sigma_g^+$) and spin
triplet ($a^3\Sigma_u^+$) configurations. The
models we adopt here were chosen because they are given in closed
analytic form with conveniently tabulated parameters.  The models 
broadly fall into two categories: (1) the Hannover polynomial
expansion (or
X-representation)~\cite{knockel2004high,salumbides2008improved}, and (2)
the Morse/Long-Range (MLR) potential~\cite{leroy2006accurate,leroy2007new}. Details regarding
these models are contained in Refs.~\cite{leroy_accurate_2009, dattani_dpf_2011,
  knoop_feshbach_2011, falke_potassium_2008, strauss_hyperfine_2010,
  baldwin2012improved}. 

\begin{figure}[t]
  \includegraphics[width=3.4in]{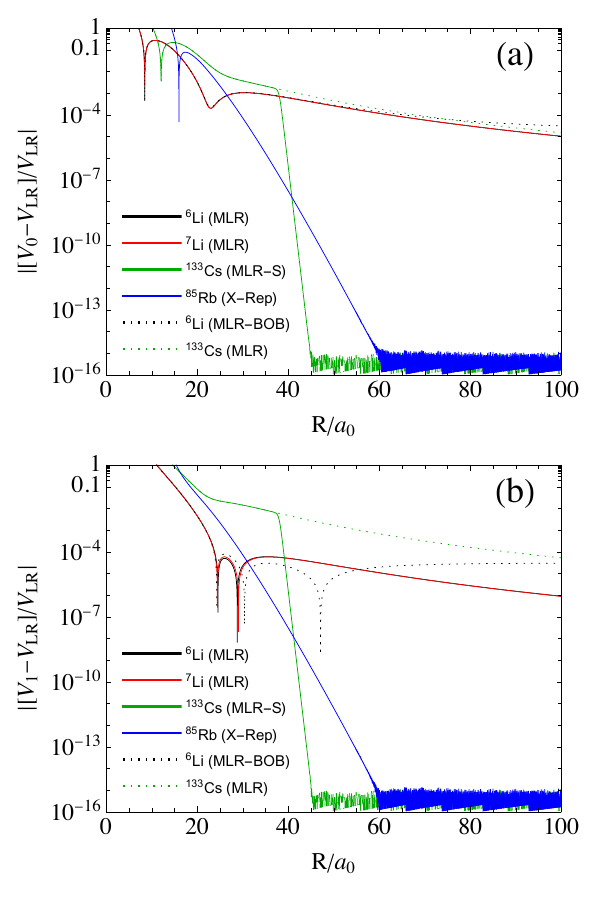}
   \caption{\label{fig:pVConv} (color online) Convergence of the singlet and
   triplet potentials to their asymptotic form is shown for a few illustrative
   cases. Panel (a) shows the singlet potentials while Panel (b) shows the triplet potentials. The solid lines correspond to potentials as they are used in this work, while the dotted curves represent unaltered potentials for $^6\text{Li}$ (dotted black) and for $^{133}\text{Cs}$ (dotted green).}
\end{figure}

The Hannover X-Rep potentials are used for $^{23}\text{Na}$~\cite{knoop_feshbach_2011},
$^{39}\text{K}$~\cite{falke_potassium_2008}, $^{40}\text{K}$~\cite{falke_potassium_2008}, $^{85}\text{Rb}$~\cite{strauss_hyperfine_2010}, and
$^{87}\text{Rb}$~\cite{strauss_hyperfine_2010}.  These potentials require essentially no modification for our
purpose; they allow for immediate and direct comparisons with experimentally observed Feshbach
resonance positions. Moreover, these potentials exhibit rapid exponential 
convergence to the asymptotic form $V_{\text{LR}}$ of
Eq.~(\ref{eq:vlr}) for $R\gtrsim 30 a_0$.  The long-range form of the
potentials in the X-representation is of the form:
\begin{equation}
  \label{eq:vex}
  V_S^{(\text{X-Rep})}(R) \rightarrow V_{\text{LR}}(R) \pm A_{\text{ex}}R^\gamma e^{-\beta_{\text{ex}} R}.
\end{equation}
When including the exponential ``exchange" term, we take the ``$+$" sign for the triplet and the ``$-$" sign for the singlet.
In Fig.~\ref{fig:pVConv}, we show the relative error of the singlet (panel (a)) and triplet (panel (b)) potentials with respect to the long-range potential $V_{\text{LR}}$ for a selection of alkali dimers. The X-Rep potential is shown only for the case of $^{85}\text{Rb}$, but other potentials of this type exhibit similar convergence.

The MLR potentials used for $^6\text{Li}$, $^7\text{Li}$ and $^{133}\text{Cs}$, on the other hand, do not behave asymptotically as Eq.~(\ref{eq:vex})~\cite{leroy_accurate_2009,dattani_dpf_2011,baldwin2012improved}.  While they do indeed converge to the form of Eq.~(\ref{eq:vlr}), that convergence is significantly slower than the exchange term, as seen by the red $^{7}\text{Li}$ curve in Fig.~\ref{fig:pVConv}. The MLR potentials for $^6\text{Li}$ also include so-called ``Born-Oppenheimer breakdown" (BOB) corrections, which are not included in the potentials for the ``reference" isotopologue $^7\text{Li}_2$.  These corrections are configuration-dependent. They behave as $~R^{-6}$ to leading order and alter the long-range potential, leading to an ``effective" $C_6$ coefficient that is different for the singlet and triplet configurations. Therefore, neither of the potentials for $^6\text{Li}$ converge to $V_{\text{LR}}$, as demonstrated by the dotted-black curves in Fig.~\ref{fig:pVConv}. Meanwhile, the MLR potentials for $^{133}\text{Cs}$ exhibit even slower convergence to $V_{\text{LR}}$, particularly for the triplet, as shown by the dotted-green curve in Fig.~\ref{fig:pVConv}. We shall soon discuss how these potentials are modified in this work in order to accommodate an MQDT treatment, which requires that the boundary condition for determining ${\bf K}_{\text{sr}}$ be applied at a separation distance where (1) the off-diagonal elements of $V_{ij}$ vanish, and (2) the diagonal elements $V_{ii}$ are reliably converged to $V_{\text{LR}}$ while and all collision channels are locally open. 

For potentials of the Hannover X-Rep type, the matching radius $R_m$ at which ${\bf K}_{\text{sr}}$ is determined may be chosen to be as small as $30 a_0$, and all quantum defects are independent of this matching radius up to about $R_m\approx 40 a_0$ where some collision channels begin to become energetically closed. For potentials of the MLR type, however, the convergence to the asymptotic form is prohibitively slow, and two options are available for improving the performance of both MQDT and FT methods.  First, one may extend the matching radius out beyond the distance at which all channels are strictly open at the risk of incurring greater energy dependence in the quantum defects.  Second, one may \emph{force} the singlet and triplet potentials to the long-range form by using a ``switching" function like Eq.~(\ref{eq:switching}).  We take the former strategy with lithium where the hyperfine-Zeeman splitting is relatively weak, and even at separation distances of $55 a_0$, the quantum defects vary smoothly, nearly linearly with energy.  For cesium, however, the higher collision channels are strongly closed beyond about $40 a_0$ and the energy dependence in ${\bf K}_{\text{sr}}$ and $\mu_S$ quickly becomes unmanagable, so we take the latter strategy and force $V^{\text{MLR}}_S\rightarrow V_{\text{LR}}$ for $R\gtrsim 40 a_0$.   

\begin{table}[b]
  \caption{\label{tbl:cn} Dispersion coefficients used in this work.}
  \begin{threeparttable}
    \begin{ruledtabular}
      \begin{tabular}{ccccc}
        dimer               & Ref.                       & $C_6 [E_h a_0^6]$  & $C_8 [E_h a_0^8]$     & $C_{10} [E_h a_0^{10}]$ \\
        \midrule
        $^6\text{Li}_2$     & \cite{tang2009nonrelativistic} & $\SI{1394.1608}{}$ & $\SI{8.3460306e+4}{}$ & $\SI{7.3744895e+6}{}$   \\
        $^7\text{Li}_2$     & \cite{tang2009nonrelativistic} & $\SI{1394.0508}{}$ & $\SI{8.3455860e+4}{}$ & $\SI{7.3741984e+6}{}$   \\
        $^{23}\text{Na}_2$  & \cite{knoop_feshbach_2011}     & $\SI{1560.0791}{}$ & $\SI{1.2496113e+5}{}$ & $\SI{8.1551411e+6}{}$   \\
        $^{39}\text{K}_2$   & \cite{falke_potassium_2008}    & $\SI{3925.9127}{}$ & $\SI{4.2237897e+5}{}$ & $\SI{4.9379591e+7}{}$   \\
        $^{40}\text{K}_2$   & \cite{falke_potassium_2008}    & $\SI{3925.9127}{}$ & $\SI{4.2237897e+5}{}$ & $\SI{4.9379591e+7}{}$   \\
        $^{85}\text{Rb}_2$  & \cite{strauss_hyperfine_2010}  & $\SI{4710.2163}{}$ & $\SI{5.7669645e+5}{}$ & $\SI{7.5912809e+7}{}$   \\
        $^{87}\text{Rb}_2$  & \cite{strauss_hyperfine_2010}  & $\SI{4710.2163}{}$ & $\SI{5.7669645e+5}{}$ & $\SI{7.5912809e+7}{}$   \\
        $^{133}\text{Cs}_2$ & \cite{baldwin2012improved}     & $\SI{6881.3838}{}$ & $\SI{1.02255e+6}{}$   & $\SI{1.5903e+8}{}$      
      \end{tabular}
    \end{ruledtabular}
  \end{threeparttable}
\end{table}

\begin{table*}[t]
  \caption{\label{tbl:scatlen} Scattering lengths for alkali dimers.}
  \begin{threeparttable}
    \begin{ruledtabular}
      \renewcommand{\arraystretch}{1.2}
      \begin{tabular}{cccccccc}
            & \multicolumn{4}{c}{Present Calculation} & \multicolumn{2}{c} {Literature}                                                     \\
         \cmidrule(lr){2-5}
         \cmidrule(lr){6-8}
         $X^1\Sigma_g^+$/$a^3\Sigma_u^+$ model         & $V_{c}^{0} [E_h/a_0^2]$     & $V_{c}^{1} [E_h/a_0^2]$ & $a_S/a_0$  & $a_T/a_0$  & $a_S/a_0$                                        & $a_T/a_0$      & Other Refs.                                  \\
        \midrule
      $^6\text{Li}_2$ (MLR)~\cite{leroy_accurate_2009,dattani_dpf_2011} & $\SI{2.65e-7}{}$                        & $\SI{1.25465e-06}{}$                & $45.166$   & $-2121.11$ &  $\SI{45.154(2)}{}$\cite{julienne2014contrasting} & $\SI{-2113(2)}{}$\cite{julienne2014contrasting}  & \cite{bartenstein2005precise, abraham1997triplet} \\
      $^7\text{Li}_2$ (MLR)~\cite{leroy_accurate_2009,dattani_dpf_2011} & $\SI{1.88e-6}{}$                        & $\SI{1.85e-6}{}$                    & $34.339$   & $-26.923$  &  $\SI{34.331(2)}{}$\cite{julienne2014contrasting} & $\SI{-26.92(7)}{}$\cite{julienne2014contrasting} &  \cite{abraham1997triplet}\\
        $^{23}\text{Na}_2$ (X-rep)~\cite{knoop_feshbach_2011}             & $0$                                     & $0$                                 & $18.820$   & $64.302$   & $\SI{18.81(80)}{}$\cite{knoop_feshbach_2011}    & $\SI{64.30(40)}{}$\cite{knoop_feshbach_2011} & \cite{samuelis_cold_2000,vanabeelen_determination_1999}  \\
        $^{39}\text{K}_2$ (X-rep)~\cite{falke_potassium_2008}             & $0$                                     & $0$                                 & $138.808$  & $-33.391$  & $\SI{138.80}{}$\cite{falke_potassium_2008}      & $\SI{-33.41}{}$\cite{falke_potassium_2008} & \cite{derrico_feshbach_2007}       \\
        $^{40}\text{K}_2$ (X-rep)~\cite{falke_potassium_2008}             & $0$                                     & $0$                                 & $104.425$  & $169.185$  & $\SI{104.42}{}$\cite{falke_potassium_2008}      & $\SI{169.18}{}$\cite{falke_potassium_2008} & \cite{derrico_feshbach_2007}        \\
        $^{85}\text{Rb}_2$ (X-rep)~\cite{strauss_hyperfine_2010}          & $0$                                     & $0$                                 & $2572.37$ & $-392.496$ & $\SI{2720}{}$\cite{strauss_hyperfine_2010}       & $\SI{-386.9}{}$\cite{strauss_hyperfine_2010} & \cite{blackley_rubidium85,roberts1998resonant}   \\
        $^{87}\text{Rb}_2$ (X-rep)~\cite{strauss_hyperfine_2010}          & $0$                                     & $0$                                 & $90.161$   & $98.867$   & $\SI{90.35}{}$\cite{strauss_hyperfine_2010}      & $\SI{99.04}{}$\cite{strauss_hyperfine_2010}  &  \cite{roberts1998resonant}  \\
        $^{133}\text{Cs}_2$ (MLR)~\cite{baldwin2012improved}              & $\SI{-2.53e-7}{}$                       & $\SI{5.9705e-7}{}$                  & $280.253$  & $2405.21$  & $\SI{280.25}{}$\cite{baldwin2012improved}        & $\SI{2405.6}{}$\cite{baldwin2012improved} & \cite{chin_precision_2004,berninger_feshbach_2013}
      \end{tabular}
    \end{ruledtabular}
  \end{threeparttable}
\end{table*}
As discussed above, the Born-Oppenheimer breakdown corrections~\cite{leroy_accurate_2009,dattani_dpf_2011} included in the MLR potentials for $^6\text{Li}$ produce different ``effective" $C_6$ coefficients for the singlet and triplet configurations.  This is undesirable for an MQDT calculation, and so we have chosen to exclude these corrections from the potential.  We have replaced the dispersion coefficients quoted in Refs.~\cite{leroy_accurate_2009,dattani_dpf_2011} with those of Ref.~\cite{tang2009nonrelativistic}, which include nonadiabatic corrections as well. For a comprehensive list of dispersion coefficients used in this work, see Table~\ref{tbl:cn}. This replacement significantly changes the singlet and triplet scattering lengths, and further changes to the potential are necessary in order to restore $a_S$ and $a_T$ to more realistic values.  A common
strategy~\cite{julienne2014contrasting, berninger_feshbach_2013} for reproducing experimental data is to adjust the volume of 
the potentials by adding a quadratic term inside the
equilibrium distance (i.e., for $R < R_{e}$ where $R_e$ is the 
potential energy minimum) of the form 
\begin{equation}
       V_{\text{shift}}(R) = V_{c}^{(S)}(R-R_{e})^2 \;\;\; \text{for}\; R<R_e.
\label{eq:potshift}
\end{equation}
Here, $V_{c}^{(S)}$ are constant parameters that may be adjusted 
to reproduce the desired scattering lengths (or particular resonance positions) and $S$ is the total spin quantum number. 

Table~\ref{tbl:scatlen} shows our scattering length calculations for all of the alkali species considered in this work. Despite the fact that the computation of single-channel scattering lengths is a relatively simple, numerically stable procedure---at least compared to solutions to large coupled channels problems---our calculations yield scattering lengths different from other published values for the same potentials.  The differences are slight, yet significant since the precise positions of magnetic Feshbach resonances are sensitive to small changes in the singlet and triplet phase shifts. These differences are discussed case-by-case in Section~\ref{results}.

\section{Multichannel Quantum Defect Theory for Ultracold Collisions}
\label{sec:mqdt-theory}

As discussed in Sec.~\ref{sec:collisionmodel}, the two-atom Hamiltonian exhibits a natural separation of energy and length scales. At short-range ($R \lesssim 30a_0$), the interaction is dominated by the deep singlet and triplet potentials, while at longer range $R\gtrsim 30a_0$, the potentials approach their comparatively weak long-range dispersion form Eq.~(\ref{eq:vlr}), offset by thresholds determined by the two-atom hyperfine-Zeeman interaction. At asymptotically large distances $R \gg \beta$, the solution may be matched to Bessel functions to determine the physical $K$-matrix.
The basic MQDT procedure is as follows: (1) Solve the Schr\"odinger equation in each of these 
three regions, the short-range region, the van der Waals, and the
asymptotic region. (2) Match the short-ranged numerical solution to the solution in the van der Waals region in order to determine the short-range $K$-matrix ${\bf K}_{sr}$, whose eigenvalues exhibit smooth, simple dependence on energy and field. (3) Match the solution in the van der Waals region where all collision channels are locally open to the appropriate asymptotic solution in order to compute a physical $K$-matrix. Here, we shall focus on steps (2) and (3) of this procedure.  For step (1), we use Johnson's log-derivative propagator~\cite{johnson1973multichannel}.

\subsection{Overview of MQDT}
Central to the implementation of MQDT, we seek a linearly independent pair of solutions, $\hat{f}_i(R)$ and $\hat{g}_i(R)$, to the single-channel Sch\"odinger equation in the presence of $V_{\text{LR}}(R)$ that are analytic in energy across the collision threshold. These reference functions satisfy
\begin{equation}
    \left ( \frac{\hbar^2}{2\mu}\left[-\frac{d^2}{dR^2}+\frac{l_i(l_i+1)}{R^2}\right]+V_{\text{LR}}(R)-E_i \right )
    \begin{Bmatrix}
      \hat{f}_i(R) \\
      \hat{g}_i(R)
    \end{Bmatrix} = 0
\label{eq:vdw-ham}
\end{equation}
with $E_i = E-E^{\text{th}}_i$. The desired reference functions are constructed
using the Milne phase amplitude method
\cite{milne1930numerical,yoo1986implementation}:  
\begin{align}
       \hat{f}_i(R)  & = \alpha_i (R) \sin{\left(\int^{R}_{R_x}{\alpha_i^{-2} (R') dR'} + \phi_i \right)} \\
       \hat{g}_i(R) & = -\alpha_i (R) \cos{\left(\int^{R}_{R_x}{\alpha_i^{-2} (R') dR'} + \phi_i \right)}
\label{eq:vdw-ref-funs}
\end{align}
where $\phi_i$ is a channel-dependent (but energy-\emph{independent}) phase and $\alpha_i(R)$ satisfies the
nonlinear differential equation
\begin{equation}
       \alpha_i(R)''+k_i^2(R) \alpha_i(R) = \alpha_i^{-3}(R).
\label{eq:alpha-fun}
\end{equation}
Here, $k_i(R) = \sqrt{2\mu[E_i-V_{\text{LR}}(R)]/\hbar^2-\ell_i(\ell_i+1)/R^2}$ is the
local wavenumber in the $i$\textsuperscript{th} channel.  It is convenient to
impose WKB-like boundary conditions~\cite{yoo1986implementation} deep in the
well (we choose $R_x=0.07 \beta$) of the long-range reference potential: 
\begin{equation}
       \alpha_i(R_x) = \frac{1}{\sqrt{k_i(R_x)}}, 
\label{eq:alpha-bc1}
\end{equation}
and
\begin{equation}
       \alpha_i'(R_x) = \frac{d}{dR}\left(\frac{1}{\sqrt{k_i(R)}}\right)_{R=R_x}.
\label{eq:alpha-bc2}
\end{equation}
The selection of the point $R_x$ in
Eqs.~\ref{eq:vdw-ref-funs}-\ref{eq:alpha-bc2} is somewhat arbitrary. All that is
required is that $V_{\text{LR}}$ is deep enough that our semi-classical boundary
conditions are reasonable.  

Fixing the energy-independent phase $\phi_i$ in Eq.~\ref{eq:vdw-ref-funs}
amounts to a ``standardization'' of the MQDT reference functions. Note that as
$R\rightarrow \infty$, $V_{\text{LR}}(R)$ in Eq.~\ref{eq:vlr} reduces to a potential of
the form $-C_6/R^6$. The strategy is to focus on the zero-energy solutions to
such a potential~\cite{burke1996multichannel,ruzic2013quantum}, 
\begin{align}
       \chi_0^{+}(R)&= \sqrt{\frac{R}{\beta}} J_{-\frac{1}{4}(2\ell+1)}\left(\frac{\beta^2}{2 R^2}\right)\\
       \chi_0^{-}(R)&= \sqrt{\frac{R}{\beta}} J_{\frac{1}{4}(2\ell+1)}\left(\frac{\beta^2}{2 R^2}\right)
\label{eq:zero-energy-funs}
\end{align}
where $J_{\nu}(x)$ is the Bessel function of the first kind. As $R\rightarrow
\infty$, $\chi_0^{+}\propto R^{\ell+1}$ and $\chi_0^{-}\propto R^{-\ell}$. One
possible standardization is to choose the standardization phase $\phi_i$ such
that $\hat{f}_i(R)$ coincides with $\chi_0^{+}(R)$ as $R \rightarrow
\infty$~\cite{burke1996multichannel}. In order to make our formulation easily
adaptable to higher partial waves, we adhere to the standardization proposed in
Ref.~\cite{ruzic2013quantum}, demanding instead that $\hat{g}(R)$ coincide with
$\chi_0^{-}$ at zero energy. There is a unique value of $\tan{\phi_i}$ that
satisfies this condition~\cite{ruzic2013quantum}, namely,  
\begin{equation}
      \tan{\phi_i} =
  -\left(\frac{W\left(\chi_0^{-},\hat{g_i}_{\phi_i=0}\right)}{W\left(\chi_0^{+},\hat{f_i}_{\phi_i=0}\right)}\right)_{R
  = R_f}
\label{eq:tanphi}
\end{equation}
where $W(x,y)$ is the Wronskian and is given by $W(x,y)=x(R) y'(R) - x'(R) y(R)$
and $R_f=20\beta$ is sufficiently large for the present study.

The linearly independent reference functions $\hat{f}_i(R)$ and $\hat{g}_i(R)$
are used to define the short-ranged $K$-matrix, ${\bf K}_{\text{sr}}$, via a boundary
condition on the general solution to Eq.~(\ref{mcse}) ${\boldsymbol{\psi}}(R)$ at $R_m$, somewhere in the van der Waals region.  We let $\hat{\bf f}(R)$ and $\hat{\bf g}(R)$ be diagonal
matrices in the field-dressed channel space with functions $\hat{f}_i(R)$ and
$\hat{g}_i(R)$, respectively, along the diagonal.  Then 
\begin{equation}
  \label{eq:srbc}
{\boldsymbol{\psi}}(R) = \hat{\bf f}(R) - \hat{\bf g}(R) {\bf K}_{\text{sr}}.
\end{equation} 
Here, $\boldsymbol{\psi}(R)$ is a matrix of solutions with elements $\psi_{i\beta}$, where $\beta$ denotes the state index, and $i$ denotes the channel component.
At very large separation distance ($R \gtrsim 20 \beta$),
$V_{\text{LR}}\rightarrow 0$ and the atomic system is described by a set of
uncoupled equations, 
\begin{equation}
    \left(\frac{\hbar^2}{2\mu}\left[-\frac{d^2}{dR^2}+\frac{\ell_i(\ell_i+1)}{R^2}\right]-E_i\right)\psi_i(R) = 0
\label{eq:vlr-ham}
\end{equation}
where $E_i = E - E_i^{\text{th}}$. For open channels with  $E_i > 0$, the
solution $\psi_i(R)$ is given by a linear combination of phase-shifted Riccati
functions which asymptotically behave as 
\begin{align}
       f_i(R) & \rightarrow \sqrt{\frac{k_i}{\pi}} \sin{\left(k_i R - l_i \frac{\pi}{2} + \eta_i\right)} \label{eq:enormf}\\
       g_i(R) & \rightarrow -\sqrt{\frac{k_i}{\pi}} \cos{\left(k_i R - l_i \frac{\pi}{2} + \eta_i\right)} \label{eq:enormg}
\end{align}
as $R \rightarrow \infty$. The parameter $\eta_i$ represents the phase that is
accumulated in the van der Waals region, and it is given by
\begin{equation}
     \tan{\eta_i}=\left(\frac{W(\hat{f}_i(R), f_i^{s}(R))}{W(\hat{f}_i(R), g^{s}_i(R))}\right)_{R=R_f}
\label{eq:eta}.
\end{equation}
Here $f_i^{s}(R)$ and $g_i^{s}(R)$ are the Riccati functions which approach
\begin{align}
       f^s_i(R) = \sqrt{\frac{k_i}{\pi}} k_i R j_\ell(kR)\rightarrow \sqrt{\frac{k_i}{\pi}} \sin{\left(k_i R - l_i \frac{\pi}{2}\right)}\\
       g^s_i(R) = \sqrt{\frac{k_i}{\pi}} k_i R n_\ell(kR) \rightarrow -\sqrt{\frac{k_i}{\pi}} \cos{\left(k_i R - l_i \frac{\pi}{2}\right)}
\label{eq:Riccati}
\end{align}
The ``energy-normalized reference functions" given in Eqs.~(\ref{eq:enormf})-(\ref{eq:enormg})
$\{f_i,g_i\}$ are related to $\{\hat{f_i},\hat{g_i}\}$ by the
following transformation,  
\begin{equation}
     \begin{pmatrix} f_i \\ g_i \end{pmatrix} = \begin{pmatrix} \mathcal{A}_i^{1/2} & 0 \\ \mathcal{A}_i^{-1/2} \mathcal{G}_i & \mathcal{A}_i^{-1/2} \end{pmatrix} \begin{pmatrix} \hat{f_i} \\ \hat{g_i} \end{pmatrix},
\end{equation}
where the parameter $\mathcal{A}_i$ is related to the energy-normalization of
$\{f_i,g_i\}$ and $\mathcal{G}_i$ accounts for the phase difference
accumulated by $\{\hat{f}_i,\hat{g}_i\}$ in $V_{\text{LR}}$
\cite{ruzic2013quantum}. These parameters are computed using the following formulas: 

\begin{equation}
    \mathcal{A}_i=-\left(\frac{W(\hat{g}_i, f_i^{s})-\tan{\eta_i}W(\hat{g}_i, g^{s}_i)}{W(\hat{f}_i, g^{s}_i)+\tan{\eta_i}W(\hat{f}_i, f^{s}_i)}\right)_{R=R_f}
\label{eq:script-A}
\end{equation}

\begin{equation}
    \mathcal{G}_i=-\left(\frac{W(\hat{g}_i, g^{s}_i)+\tan{\eta_i}W(\hat{g}_i, f^{s}_i)}{W(\hat{f}_i, g^{s}_i)+\tan{\eta_i}W(\hat{f}_i, f^{s}_i)}\right)_{R=R_f}
\label{eq:script-G}
\end{equation}

For closed channels ($E_i < 0$), the solution is a superposition of
$\hat{f}_i(R)$ and $\hat{g}_i(R)$ that vanishes as $R\rightarrow \infty$: 
\begin{equation}
     \hat{f_i}(R)+\cot{\gamma_i}\hat{g_i}(R) \rightarrow \; \propto e^{-\kappa_i R}.
\end{equation}
Here, $\kappa_i = \sqrt{2\mu \abs{E_i}/\hbar^2}$ and $\gamma_i$ is a parameter that
determines what combination of $\{\hat{f}_i,\hat{g}_i\}$ vanishes as $R
\rightarrow \infty$. It is computed by
\begin{equation}
     \tan{\gamma_i}=\left(\frac{W(e^{-\kappa_i r}, \hat{g}_i(R))}{W(e^{-\kappa_i r}, \hat{f}_i(R))}\right)_{R=R_f}
\label{eq:gamma}
\end{equation}

With the energy-dependent MQDT parameters $\mathcal{A}$, $\mathcal{G}$, and
$\cot{\gamma}$ in hand, one may determine the $K$-matrix defining the
asymptotic boundary condition with respect to functions $f_i(R)$ and $g_i(R)$,
namely $\boldsymbol{\psi}(R)\rightarrow \mathbf{f}-\mathbf{K} \mathbf{g}$. First, ${\mathbf K}_{\text{sr}}$ is partitioned into blocks depending on which channels are asymptotically opened ($P$) or closed ($Q$):
\begin{equation}
    {\mathbf K}_{\text{sr}} =
    \begin{pmatrix}
    K^{sr}_{PP} & K^{sr}_{PQ} \\
    K^{sr}_{QP} & K^{sr}_{QQ} 
    \end{pmatrix}
    \label{eq:Ksr-partition}.
\end{equation}
Then, we use the closed-channel parameter $\gamma$ to transform the $N \times N$ short-range reaction matrix into an $N_P \times N_P$ matrix using the channel-closing formula
\begin{equation}
  \Tilde{\bf K} = {\bf K}^{\text{sr}}_{PP}-{\bf
    K}^{\text{sr}}_{PQ}({\bf K}^{\text{sr}}_{QQ} + {\mathbf \cot{\boldsymbol{\gamma}}} )^{-1}{\bf K}^{\text{sr}}_{QP}.
\end{equation}
This transformation accounts for the reflected amplitude arising from closed channels, and captures the physics of closed-channel resonances. Next, $\Tilde{\bf K}$ must be properly normalized with respect to energy. This is accomplished with the expression,
\begin{equation}
    {\bf  K} = {\bf \mathcal{A}}^{1/2}\Tilde{\bf K}({\bf 1}+{\bf
      \mathcal{G}}\Tilde{\bf K})^{-1}{\bf \mathcal{A}}^{1/2}.
\end{equation}
 This ${\bf K}$, however, is not yet the full physical $K$-matrix because it does not include effects from the additional phase $\eta$ which $\{f,g\}$ acquire with respect to $\{f^{s},g^{s}\}$. We obtain a physical $S$-matrix by:
\begin{equation}
{\bf S^{\text{phys}}}= e^{i\eta}\frac{{\bf 1} + i{\bf K}}{{\bf 1} - i{\bf K}}e^{i\eta}
\end{equation}
from which ${\bf K}^{\text{phys}}$ is obtained by
\begin{equation}
    {\bf K}^{\text{phys}} = i \frac{{\bf 1} - {\bf S}^{\text{phys}}}{{\bf 1} + {\bf S}^{\text{phys}}}.
\end{equation}
Note that in the
 above expressions, $\boldsymbol{\gamma}$, $\bf{\mathcal{A}}$,
 $\bf{\mathcal{G}}$, and $\boldsymbol{\eta}$ are diagonal matrices of the
 corresponding MQDT functions evaluated at the appropriate channel energy
 $E-E^{\text{th}}_i$.

For ultracold collisions in the lowest channel, ${\bf\tilde{K}}$, ${\bf K}$ and ${\bf K}^{\text{phys}}$ are each reduced to a single matrix element, and one can write the
physical $K$-matrix element as
\begin{equation}
    {K^{\text{phys}}}= \frac{\tan{\eta}+{K}}{1-{K}\tan{\eta}},
\end{equation}
or in terms of the $s$-wave phase shift as:
\begin{equation}
K^{\text{phys}}=\tan{\delta} 
\end{equation}
We are primarily interested in the scattering length $a$, which is related to
the $s$-wave phase shift $\delta$ by
\begin{equation}
  a = - \lim_{k \rightarrow 0}{\frac{\tan{\delta}}{k}}
\end{equation}
In all calculations presented here, we compute the MQDT functions using
Eqs.~\ref{eq:eta}, \ref{eq:script-A}, \ref{eq:script-G}, and \ref{eq:gamma}.

\subsection{Frame Transformation}
At short separation distances, i.e., $R \lesssim 30 \; a_0$, the physics is
dominated by the deep Born-Oppenheimer potentials $V_S(R)$. Therefore, to a good approximation, any hyperfine
or Zeeman interactions can be neglected at short range, and the atomic system
can be described by a set of uncoupled equations in the singlet and triplet channels written here only for the $s$-wave,  
\begin{equation}
\label{eq:TISE}
    \left(-\frac{\hbar^2}{2\mu}\frac{d^2}{dR^2} + V_S(R) - E \right)\psi_S(R) = 0
\end{equation}
\begin{figure*}[t]
  \includegraphics[width=6.8in]{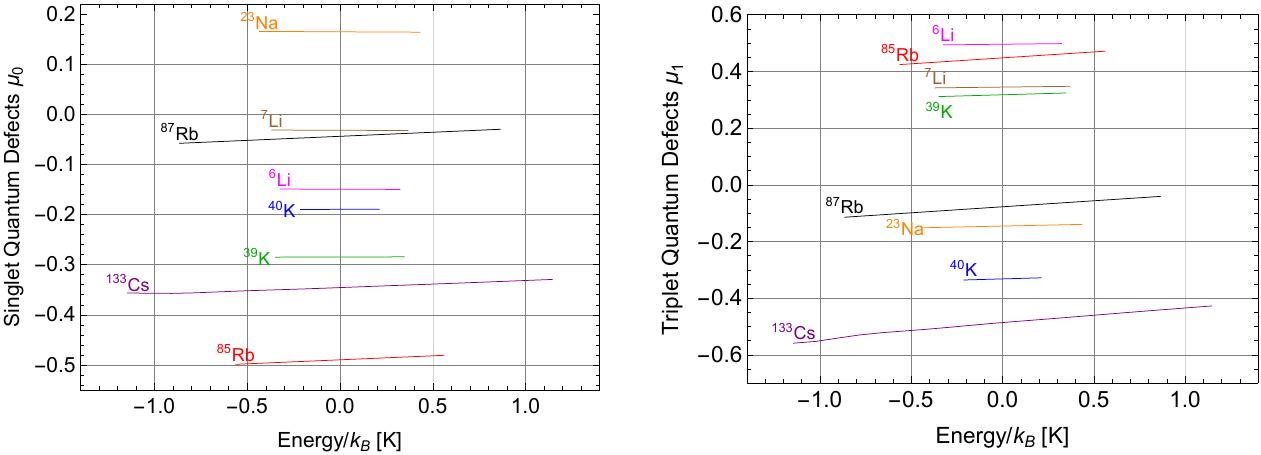}
  \caption{\label{fig:QD} (color online) Quantum defects for both the singlet (left) and triplet
    (right) potential energy functions are shown for each homonuclear dimer.
    The energy (divided by the Boltzmann constant, $k_B$) scale is given in kelvin.  The range of energies for each case is fixed by the maximum
    separation of collision thresholds, ranging from $-|\Delta E^{\text{th}}|$ to $+|\Delta E^{\text{th}}|$, where
    $\Delta E^{\text{th}} = \max{E^{\text{th}}_i}-\min{E_i^{\text{th}}}$ at the largest
    fields considered ($1200$ G). }
  \end{figure*}
In Figure~\ref{fig:QD} we show the quantum defects $\mu_S(E)$ in the singlet and triplet eigenchannels as a function of energy.  The range of energy is different for each dimer, and is determined by the maximum energy difference between collision thresholds $\Delta E^{\text{th}} = \max{E^{\text{th}}_i}-\min{E_i^{\text{th}}}$ at the largest fields considered, $1200$G. For collisions at energies near the lowest threshold ($\min{E_i^{\text{th}}}$), there may be a resonance due to a bound state attached to the highest threshold ($\max{E_i^{\text{th}}}$), so the lowest energy one may imagine evaluating the quantum defect at is $-\Delta E^{\text{th}}$.  Meanwhile, one may imagine collisions of atoms prepared in excited magnetic levels undergoing collisions that occur at relatively high energy ($\Delta E^{\text{th}}$) with respect to the ground-state channel.  These collisions are not explicitly considered in this work, but nevertheless suggest that one requires the quantum defects, in principle, over a range $-|\Delta E^{\text{th}}| < E < |\Delta E^{\text{th}} |$. More technically, the scale $\Delta E^{\text{th}}$ is set by EDFT calculation, which requires we compute the channel-weighted average energy Eq.~(\ref{eq:averageE}).
As one varies atomic mass from $^{6}\text{Li}$ to $^{133}\text{Cs}$, one finds that $\Delta E^{\text{th}}$ varies also from $\sim 0.25\text{K} \approx 5\text{GHz}$ to about $\sim 1\text{K} \approx 20\text{GHz}$.  Over the scale of relevant energies, the defects themselves vary roughly linearly, with slopes tending roughly to increase with mass.

To find the solution $\psi_S(R)$, we numerically integrate
Eq.~(\ref{eq:TISE}) from $R\approx a_0$,
sufficiently small so that $\psi(R) \rightarrow 0$ due to the hard repulsive core of the potential, out to $R_m\sim 40 \; a_0$ (or $55a_0$ for lithium as described in Section~\ref{subsec:STPots}). Then we match each solution to a linear
combination of $\hat{f}(R)$ and $\hat{g}(R)$ at $R_m$ to determine the singlet
and triplet quantum defects $\mu_{S}$  at zero energy. These \emph{single-channel} quantum defects are determined by imposing a single channel boundary condition (analogous to Eq.~(\ref{eq:srbc}) on the numerical solution $\psi_S(R)$ to Eq.~(\ref{eq:TISE}),
\begin{equation}
    \psi_S(R)\rightarrow \hat{f}(R)-\hat{g}(R)\tan{(\pi \mu_S)}.
\end{equation}
At large $R$, however, the total electronic spin $S$ is no longer a good
quantum number and the interaction between the particles is no longer diagonal
in the molecular basis $\ket{\lambda}=\ket{S M_S I M_I}$. The 
frame-transformation provides a powerful approximation to the short-range reaction matrix ${\bf K}^{\text{sr}}$ in the basis $\ket{i}$ that defines the collision channels in which the system is diagonal at large $R$
\cite{burke1996multichannel}:
\begin{equation}
     K^{\text{sr}}_{i,i'} = \sum_{\lambda}{\bra{i}\ket{\lambda}}\tan{(\pi \mu_{S}(E))}\bra{\lambda}\ket{i'}.
     \label{eq:Ksr-hf}
\end{equation}
 In absence of an external magnetic field, asymptotic channels are simply the the properly symmetrized hyperfine states Eq.~(\ref{eq:sym-basis}). However, if
there is an applied field, the asymptotic dissociation channels are now eigenstates of the full $\mathbf{H}^{\text{HZ}}$, as in Eq.~(\ref{eq:fielddressing}).

The short-ranged reaction matrix obtained from Eq.~(\ref{eq:Ksr-hf}) 
depends only on the single-channel quantum defects $\mu_S(E)$ and the field-dependent transformation that accomplishes the dressing. It is not entirely clear, however, at what energy $E$ one should evaluate the quantum defects when computing $K^{\text{sr}}$ from Eq.~(\ref{eq:Ksr-hf}), for the defects themselves are functions of energy measured with respect to the common singlet and triplet thresholds (zero), while the collision energy is measured with respect to the asymptotic threshold energies $E^{\text{th}}_i$ computed in Eq.~(\ref{eq:Ethresh}).

As a first approximation, one may assume that the energy dependence of the quantum defects is negligible, and simply evaluate $\mu_S$ at zero energy.  This results in what we call the \emph{energy independent frame transformation} (EIFT). 

A better approximation, which results in the \emph{energy dependent frame transformation} (EDFT), is to evaluate $\mu_S(E)$ at the channel-weighted average energy~\cite{burke1996multichannel}
\begin{equation}
\label{eq:averageE}
  \bar{E}_\lambda = \sum_{i}{(E-E^{\text{th}}_i)\lpipe \braket{\lambda}{i} \rpipe^2}.
\end{equation}

Both of the EIFT and EDFT approximations circumvent the need to solve a set of coupled equations, allowing all scattering observables to be computed with \emph{single} channel calculations only. The more rigorous boundary condition Eq.~(\ref{eq:srbc}) needed for MQDT, on the other hand, requires a CC calculation in the region $R \le R_m$. We refer to calculations that stem from Eq.~(\ref{eq:srbc}) as ``full" MQDT calculations, labeled as ``MQDT" in figures that follow.
\section{Results}
\label{results}
Here, we present results for the scattering length versus magnetic field for
homonuclear collisions of alkali atoms ranging from lithium to cesium. We
focus on $s$-wave collisions only, and identify the positions magnetic Feshbach
resonances and zero crossings in the lowest collision channel for a given $M_F$
block. These are compiled in Table~\ref{tbl:FieldFeatures} along with available experimental data. Empty cells in the last column indicate that we were unable to find an experimental measurement in the literature. The locations of many zeros associated with narrow resonances are difficult to observe experimentally, and several high-field resonances have yet to appear in the literature. All calculations are performed at a collision energy of $1\mu\text{K}$.

\begin{table*}[!t]
  \caption{\label{tbl:FieldFeatures} Features in the field-dependent $s$-wave
    scattering length for all alkali species considered here.  Calculations are performed at a collision energy of $1 \mu \text{K}$. Results are in gauss (G).}
  \begin{threeparttable}
    \begin{ruledtabular}
      \begin{tabular}{ccccccc}
        Atom                                         & Feature & CC         & MQDT       & EDFT       & EIFT       & EXPERIMENT                                                                               \\
        \midrule
        \multirow{4}{*}{$^{6}\text{Li}$, $M_F=0$}    & zero    & $527.407$  & $527.220$  & $527.200$  & $526.851$  & $527.5(2)$ ~\cite{du2008observation}, $528(4)$~\cite{ohara_measurement_2002}, $530(3)$~\cite{jochim2002magnetic}                                               \\
                                                     & pole    & $543.286$  & $543.284$  & $543.282$  & $542.934$  & $543.28(1)$~\cite{schunck2005feshbach}, $543.286(3)$~\cite{hazlett2012realization} \\
                                                     & zero    & $543.387$  & $543.384$  & $543.382$  & $543.034$  &                                                                      \\
                                                     & pole    & $832.180$   & $832.186$   & $831.779$   & $831.527$   & $834.1(1.5)$~\cite{bartenstein2005precise}, $822(3)$~\cite{zwierlein2004condensation}                                         \\
                                                     &         &            &            &            &            &                                                                                    \\
        \multirow{3}{*}{$^{7}\text{Li}$, $M_F=+2$}   & zero    & $140.909$  & $140.917$  & $139.854$  & $139.614$  &                                                                      \\
                                                     & zero    & $543.438$  & $543.435$  & $544.420$  & $543.573$  & $543.6(1)$~\cite{pollack2009extreme}                                               \\
                                                     & pole    & $737.716$  & $737.717$  & $737.949$  & $736.334$  & $737.69(12)$~\cite{dyke2013finite}                                                 \\
                                                     &         &            &            &            &            &                                                                                    \\
        \multirow{4}{*}{$^{23}\text{Na}$, $M_F=+2$}  & pole    & $851.074$  & $851.073$  & $852.215$  & $865.066$  & $851.0(2)$~\cite{knoop_feshbach_2011}                                              \\
                                                     & zero    & $851.083$  & $851.083$  & $852.225$  & $865.076$  &                                                                      \\
                                                     & pole    & $905.149$  & $905.147$  & $905.159$  & $917.777$  & $905.1(4)$~\cite{knoop_feshbach_2011}                                              \\
                                                     & zero    & $906.193$  & $906.191$  & $906.203$  & $918.780$  &                                                                      \\
                                                     &         &            &            &            &            &                                                                                    \\
        \multirow{8}{*}{$^{39}\text{K}$, $M_F=+2$}   & zero    & $25.427$   & $25.424$   & $25.343$   & $25.764$   &                                                                      \\
                                                     & pole    & $25.886$   & $25.889$   & $25.836$   & $26.236$   & $25.85(10)$~\cite{derrico2007feshbach}                                             \\
                                                     & zero    & $350.374$  & $350.364$  & $350.492$  & $350.720$  & $350$~\cite{fattori_potassium39}, $350.4$~\cite{derrico2007feshbach}                                                                     \\
                                                     & pole    & $402.461$  & $402.462$  & $402.338$  & $402.558$  & $403.4(7)$~\cite{derrico2007feshbach}                                              \\
                                                     & zero    & $741.931$  & $744.930$  & $745.000$  & $750.397$  &                                                                     \\
                                                     & pole    & $744.936$  & $744.935$  & $745.005$  & $750.402$  &                                                                     \\
                                                     & zero    & $751.886$  & $751.882$  & $751.935$  & $757.268$ &                                                                      \\
                                                     & pole    & $752.277$  & $752.280$  & $752.334$  & $757.65$   & $752.3(1)$~\cite{derrico2007feshbach}                                              \\
                                                     &         &            &            &            &            &                                                                                    \\
        \multirow{4}{*}{$^{40}\text{K}$, $M_F=-7$}   & pole    & $12.661$   & $12.661$   & $13.009$   & $14.557$   &                                                                      \\
                                                     & zero    & $12.663$   & $12.663$   & $13.0132$   & $14.558$   &                                           \\
                                                     & pole    & $224.222$  & $224.222$  & $223.758$  & $223.909$  & $224.2(1)$ \cite{regal2003measurement}                                                                    \\
                                                     & zero    & $231.432$  & $231.432$  & $231.151$  & $231.287$  & $233.9(1)$ \cite{regal2003measurement}                                             \\
                                                     &         &            &            &            &            &                                                                                    \\
        \multirow{4}{*}{$^{85}\text{Rb}$, $M_F=+4$}  & zero    & $850.572$  & $850.571$  & $847.973$  & $868.110$   &                                                                      \\
                                                     & pole    & $851.755$  & $851.755$  & $850.911$  & $870.204$  & $852.3(3)$~\cite{blackley_rubidium85}                                                                                   \\
                                                     & zero    & $1068.352$  & $1068.352$  & $1070.585$  & $1087.537$  &                                                                                    \\
                                                     & pole    & $1070.787$  & $1070.787$  & $1073.679$  & $1092.366$  &                                                                                    \\
                                                     &         &            &            &            &            &                                                                                    \\
        \multirow{8}{*}{$^{87}\text{Rb}$, $M_F=+2$}  & pole    & $406.883$  & $406.883$  & $400.758$  & $446.639$  & $406.2(3)$~\cite{marte_rubidium87}                                                                     \\
                                                     & zero    & $406.884$  & $406.884$  & $401.249$  & $446.897$  &                                                                     \\
                                                     & pole    & $686.396$  & $686.396$  & $692.704$  & $753.618$  & $685.4(3)$~\cite{marte_rubidium87}                                                                    \\
                                                     & zero    & $686.403$  & $686.402$  & $692.706$  & $753.941$  &                                                                     \\
                                                     & pole    & $911.651$  & $911.651$  & $933.705$  & $1008.810$  & $911.7(4)$~\cite{marte_rubidium87}                                                                      \\
                                                     & zero    & $911.652$  & $911.652$  & $933.707$  & $1008.840$  &                                                                      \\
                                                     & pole    & $1007.71$  & $1007.710$  & $986.280$  & $1046.440$  & $1007.40(4)$~\cite{volz_rubidium87}, $1007.3(4)$~\cite{marte_rubidium87}                                                                     \\
                                                     & zero    & $1007.91$  & $1007.910$  & $986.835$  & $1046.780$  & $1007.60(3)$~\cite{volz_rubidium87}                                                                       \\
                                                     &         &            &            &            &            &                                                                                    \\
        \multirow{6}{*}{$^{133}\text{Cs}$, $M_F=+6$} & pole    & $-8.654$ & $-9.693$ & $-56.018$ & $59.594$  &  $-11.7$~\cite{chin_precision_2004}                                                                               \\
                                                     & zero    & $10.155$  & $10.139$  & $4.308$  & $86.218$  & $17.26(20)$\cite{mevznarvsivc2019cesium}, $17.119(2)$\cite{gustavsson_cesium}     \\
                                                     & pole    & $545.846$  & $545.866$  & $468.879$  & $599.953$  & $549$~\cite{berninger_feshbach_2013}                                                                                   \\
                                                     & zero    & $551.406$  & $551.410$   & $538.332$  & $616.613$  & $553.73(2)$~\cite{berninger_feshbach_2013}                                                                                   \\
                                                     & pole    & $822.933$  & $823.140$   & $743.691$  & $803.191$  & $787$\cite{berninger_feshbach_2013}                                                                                 \\
                                                     & zero    & $901.203$  & $901.202$  & $896.486$  & $933.785$  &                                                                      \\
      \end{tabular}
    \end{ruledtabular}
  \end{threeparttable}
\end{table*}

\subsection{Lithium} 
\begin{figure*}[t]
  \centering
  \includegraphics[width=6.8in]{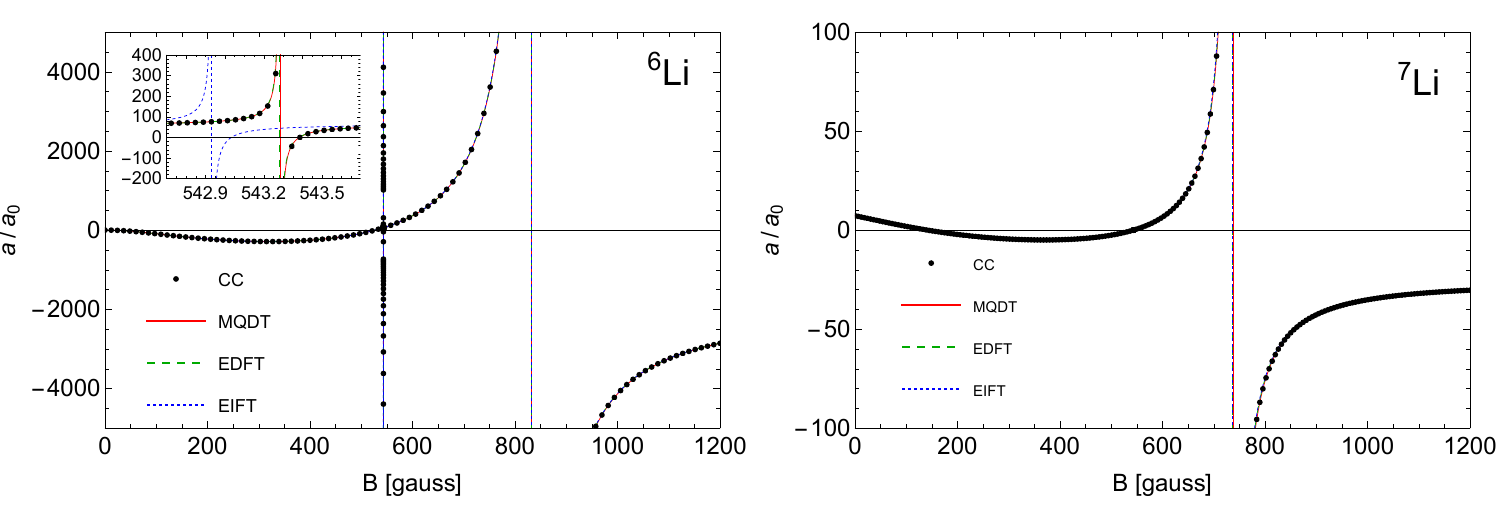}
   \caption{\label{fig:Li7} (color online) The field-dependent
     scattering length for $^6$Li collisions with $M_F =0$ and $^7$Li
     collisions in the ground state with $M_F=2$ at collision energy
     $1\mu\text{K}$ is shown. The inset on the $^6$Li plot shows the
     narrow resonance at $543.286$G. }
\end{figure*}

For both isotopes of lithium, we use the MLR potentials of
Ref.~\cite{leroy_accurate_2009,dattani_dpf_2011}, but with two significant
alterations.  First, we use the dispersion coefficients $C_n$ tabulated in
Ref.~\cite{tang2009nonrelativistic}, which include effects arising from the
finite mass of the atomic nuclei. Second, we modify the short-ranged behavior of the singlet/triplet potentials by adding a term of the form Eq.~(\ref{eq:potshift}). 

The first alteration is particularly necessary for our purpose of developing and testing the accuracy of MQDT methods because the dispersion coefficients reported in Refs.~\cite{leroy_accurate_2009,dattani_dpf_2011} differ for the singlet and triplet channels. It is desirable to have the \emph{same} long-range behavior in each collision channel for MQDT, so that the MQDT parameters $\mathcal{G(E), A(E)}, \eta(E), \gamma(E)$ can be computed uniquely.  Having modified the dispersion coefficients, it is essential to include the short-ranged potential in Eq.~(\ref{eq:potshift}) in order to restore the singlet and triplet scattering lengths to physically realistic values.

References~\cite{leroy_accurate_2009,dattani_dpf_2011} report $^{6}\text{Li}$ scattering lengths $a_S=45.05(9) a_0$ and $a_T=-3602(95) a_0$, where the reported uncertainties arise from statistical errors in the direct potential fit. Our scattering length calculations using their subroutines~\cite{leroycodes} for generating the potentials ``out of the box" yield $a_S/a_0 = 45.046$ and $a_T/a_0 = -3430.2$, showing excellent agreement for the singlet $a_S$, but $\approx 5\%$ discrepancy in the triplet $a_T$, for which we cannot account. We have conducted rigorous tests of our calculations, as detailed in Appendix~\ref{RawitcherTest}. The \emph{same} code used to solve the coupled-channels problem was used to compute the singlet and triplet scattering lengths.

We adjust the parameters $V_{c}^{(S)}$ primarily to reproduce the scattering lengths reported in Ref.~\cite{julienne2014contrasting}. Ultimately, our reported scattering lengths for $^6\text{Li}$ in Table~\ref{tbl:scatlen} differ slightly from Ref.~\cite{julienne2014contrasting} because we have made additional adjustments to match the position for the narrow $s$-wave resonance near $543$G to the experimental observation of Hazlett et al.~\cite{hazlett2012realization}. 

For $^{6}\text{Li}$ we consider elastic collisions in the lowest channel with $M_F = 0$. Table~\ref{tbl:FieldFeatures} lists the zeroes and poles of the $s$-wave scattering length as determined by the coupled-channels (CC) calculation, MQDT, EDFT, and EIFT. The left graph in Fig.~\ref{fig:Li7} plots $a(B)$ for field values ranging from $0$G to $1200$G. There is broad resonance near $832$G and a narrow feature around $543$G, which is shown more clearly in the inset. Both MQDT and EDFT come within $1$mG of the CC calculation for this narrow resonance, while EIFT is off by $0.3$G. However, both EDFT and EIFT slightly underestimate the location of the broad resonance at $832.18$G, whereas MQDT almost exactly agrees with the CC calculation.  

Several groups have experimentally determined the resonance features of this collision~\cite{jochim2002magnetic, ohara_measurement_2002,du2008observation,strecker2003conversion,schunck2005feshbach,hazlett2012realization,zwierlein2004condensation,bartenstein2005precise}. Measurements made by Jochim et al.~\cite{jochim2002magnetic} and O'Hara et al.~\cite{ohara_measurement_2002} in 2002 place the location of the first zero-crossing at $530\pm(3)$G and $528 \pm 4$G, respectively.
In 2008, Du et al. more accurately determined the position to be $527.5\pm 0.2$G~\cite{du2008observation}. Our CC calculation agrees with the latter value, while MQDT and EDFT fall just outside the experimental uncertainty. The location of the narrow resonance has been measured by Refs.~\cite{strecker2003conversion,schunck2005feshbach}. However, to date, Hazlett et al.~\cite{hazlett2012realization} has made the most precise measurement at $543.286(3)$ G. We pin our model for the potential curves so that this resonance position is to reproduce by our CC calculations to better than $1$mG. The results of MQDT and EDFT are nearly within the error bars of this observation. Using RF spectroscopy on weakly bound molecules, Bartenstein et al.~\cite{bartenstein2005precise} measured the position of the wide resonance to be $834.1(1.5)$G, which our CC and MQDT calculations fall just short of.

Scattering length calculations using the potentials of 
Refs.~\cite{leroy_accurate_2009,dattani_dpf_2011} for 
the case of $^{7}\text{Li}$ show reasonable, but not perfect agreement with values reported in those papers.
Refs~\cite{leroy_accurate_2009,dattani_dpf_2011} find $a_S/a_0 = 34.22(9)$ and
$a_T/a_0 = -27.80(2)$, while we find $a_S/a_0 = 34.222$ and $a_T/a_0 = -27.891$.
As with $^{6}\text{Li}$, we replace the dispersion coefficients of Refs.~\cite{leroy_accurate_2009,dattani_dpf_2011} with those of Ref.~\cite{tang2009nonrelativistic}. Having done so, the parameters $V_{c}^{(S)}$ of Eq.~(\ref{eq:potshift}) are adjusted to give the best agreement possible with experimental measurements of the scattering length node near $544$G and the wide resonance near $738$G.  This yields scattering lengths comparable to those reported in Ref.~\cite{julienne2014contrasting}.

The right plot in Fig.~\ref{fig:Li7} shows the field-dependent scattering length for $^{7}\text{Li}$ elastic collisions in the ground state with $M_F=2$. Just as in the $^{6}\text{Li}$ case, we find that MQDT is nearly in perfect agreement with the CC calculation, only underestimating the zero crossings in $a(B)$ by a few mG. Both EDFT and EIFT do slightly worse, coming within $1$G of the full coupled-channels calculation. Pollack et al.~\cite{pollack2009extreme} observed that the scattering length passes through a zero crossing at
$B_0 = 543.6(1)$G with a slope of $\Delta a/\Delta B = 0.08 a_0/\text{G}$. They fit the resonance peak to $736.8(2)$G. Our best fit (performed manually) yields positions shown in Table~\ref{tbl:FieldFeatures}. Our CC and MQDT calculations are nearly within the error bars of this observation and our calculation of the slope of $a(B)$ is $\Delta a / \Delta B=0.079 a_0/\text{gauss}$, in agreement with their observations.
\subsection{Sodium}
\begin{figure}[t]
  \centering
  \includegraphics[width=\linewidth]{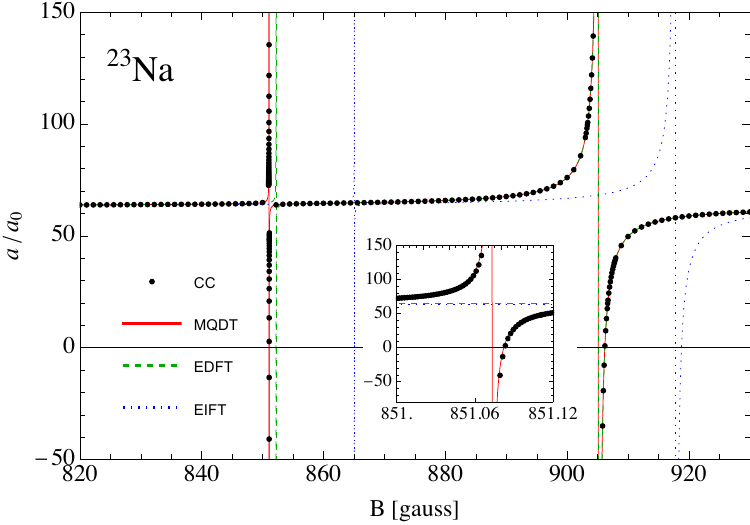}
  \caption{\label{fig:Na23} (color online) The scattering length (in units of $a_0$) for $^{23}\text{Na}$
    collisions in the lowest channel with total $M_F=2$ at threshold is shown as a function of magnetic field.}
\end{figure}

Early experiments~\cite{inouye1998observation,stenger1999strongly} reported two s-wave resonances for sodium atoms prepared in the $\ket{f,m_f}=\ket{1,1}$ hyperfine state (i.e. ground state of the $M_F=+2$ block), one near $853$G and another narrower one near $907$G. The accuracy of those measurements was limited by magnetic field stability to about $20$G. Later experiments~\cite{knoop_feshbach_2011} greatly improved the accuracy of these resonance positions, made additional measurements of higher partial wave resonances, and developed improved singlet and triplet potential energy functions of the Hannover form.

We adopt the sodium potential energy function developed in Ref.~\cite{knoop_feshbach_2011} without modification. Field values for zeroes and pole positions in the scattering length are tabulated in Table~\ref{tbl:FieldFeatures}. In Fig.~\ref{fig:Na23}, we plot the scattering length from $820$G to $930$G, showing the two $s$-wave resonances in this range.  We found no other $s$-wave resonances with width greater than approximately $1$mG for fields less than $1200$G, but there is another narrow resonance at very high field (not shown) near $2055$G.

Comparing the MQDT, EDFT, and EIFT with converged CC calculations reveals that while the EIFT is able to reproduce the qualitative features of $a(B)$, the positions of resonances are consistently overestimated by about $14$G. Improvements afforded by the EDFT are significant.  The position of the narrow resonance near $851$G is only overestimated by about $1$G, while the position of the wide resonance near $905$G is overestimated by only $10$mG. Moreover, our CC and MQDT calculations are in agreement with Ref.~\cite{knoop_feshbach_2011}, which reports values of $851.0(2)$G and $905.1(4)$G for the two resonances.

 \subsection{Potassium}
\begin{figure*}[t]
   \centering
   \includegraphics[width=7in]{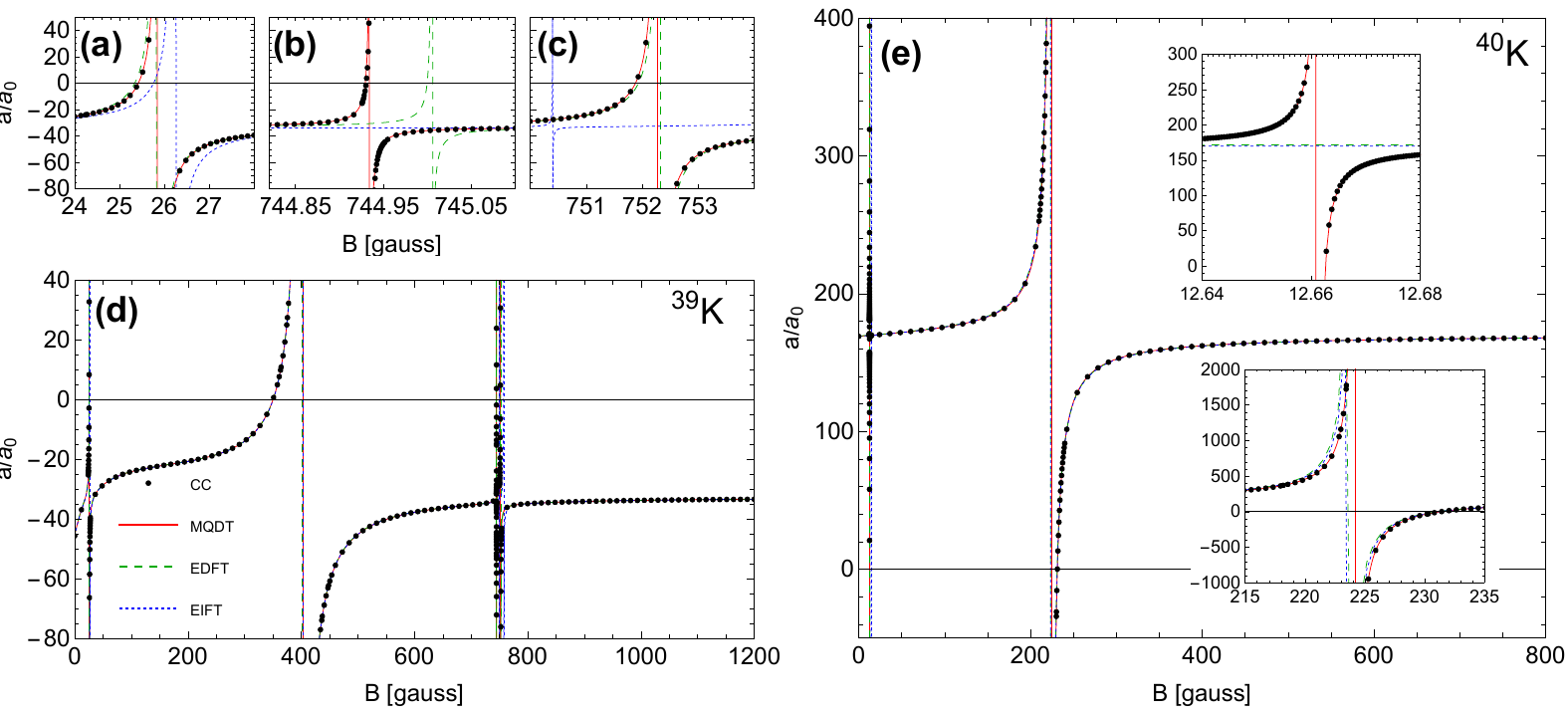}
   \caption{\label{fig:K40} (color online) (d) The scattering length (in units of $a_0$) for $^{39}\text{K}$ in the lowest channel with total $M_F = 2$, with (a)-(c) showing the narrow resonances at $25.886$G, $744.936$G, and $752.277$G, respectively. (e) The scattering length (in units of $a_0$) for $^{40}\text{K}$
     collisions in the lowest channel with total $M_F=-7$ at threshold is shown as a function of magnetic field. Insets in the $^{40}\text{K}$ plot show the resonances near $12$G and $224$G in greater detail.}
 \end{figure*}

 We use the potassium potential energy functions of Ref.~\cite{falke_potassium_2008} without modification. For $^{39}\text{K}$, we consider elastic collisions in the lowest channel with $M_F=2$. The locations of poles and zero crossings in the $s$-wave scattering length are provided in Table~\ref{tbl:FieldFeatures} and Fig.~\ref{fig:K40}(d) plots $a(B)$ for fields ranging from $0$G to $1200$G. Figs.~\ref{fig:K40}(a)-(c) show the three narrow resonances at $25.886$G, $744.936$G, and $752.277$G in more detail. Comparing MQDT, EDFT, and EIFT with the full coupled channels calculation, we find that all three methods are able to reproduce the broad resonance near $402.461$G and the narrow resonance at $25.886$G (as determined by our CC calcualtions) to within $1$G. However, EIFT fares worse for the two resonances at higher fields, overestimating the locations of poles and zeroes by about $5$G while both MQDT and EDFT are within $0.1$G of the CC calculation. Note that the sharp resonance in the EIFT calculation in Panel (c) of Fig.~\ref{fig:K40} is the same feature shown in Panel (b) for the other three calculations.

 D'Errico et al~\cite{derrico2007feshbach} found resonances in a
 number of channels.  In the $M_F=2$ block, they measured resonances at
 $25.85(10)$G, $403.4(7)$G, and $752.3(1)$G, which are nearly in agreement with our CC and MQDT calculations. However, they missed a predicted narrow near $745$G. Chapurin et al~\cite{chapurin2019precision} have recently made a
 precise measurement of a low-field resonance in $^{39}\text{K}$ in
 the $M_F = -2$ block, finding a resonance position $33.5820(14)$G, which represents a significant improvement over an earlier measurement~\cite{roy2013test}. Our CC calculations using the unmodified potential functions of Ref.~\cite{falke_potassium_2008} yield a resonance position of $33.5780$G. 

For $^{40}\text{K}$, we consider elastic collisions in the lowest channel with $M_F=-7$. The scattering length for magnetic fields between $0$G and $800$G is shown in Fig.~\ref{fig:K40}(e). There is a very narrow resonance near $12.66$G (which is shown in greater detail in the top inset) and a broader feature near $224$G (shown in the bottom inset). Analyzing our results, we find that MQDT and CC calculations exactly agree on the location of every pole and zero crossing in $a(B)$. EDFT and EIFT are slightly less accurate, differing from the coupled-channels calculation by $\sim 1$G. Looking at available experimental data, Ref.~\cite{regal2003measurement} determined the position of the broad resonance to be $224.21\pm 0.05$G, with a width $\Delta = 9.7 \pm 0.6$G, which nearly agrees with our CC calculation. 

 \subsection{Rubidium}
\begin{figure*}[t]
   \centering
   \includegraphics[width=7in]{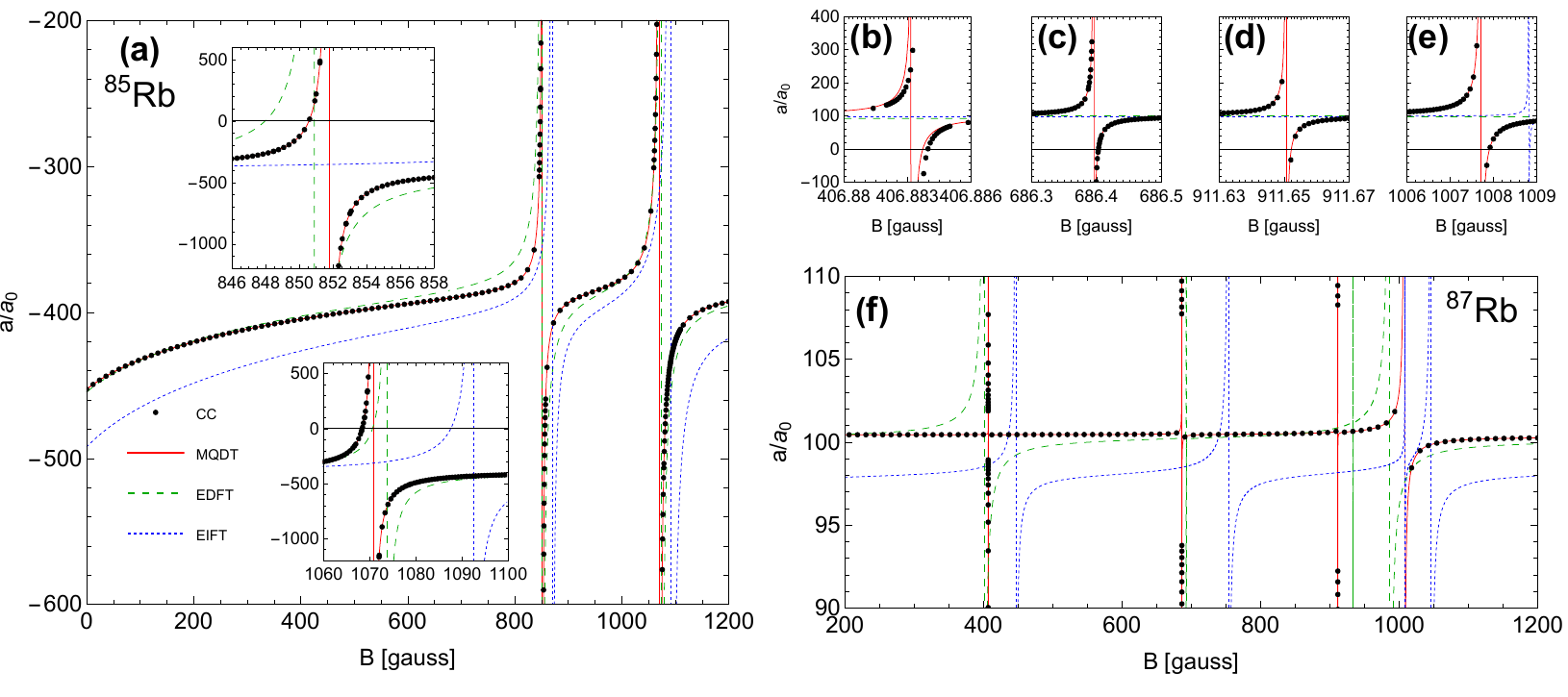}
   \caption{\label{fig:Rb87} (color online)  (a) The scattering length (in units of $a_0$) for $^{85}\text{Rb}$
     collisons in the lowest channel with total $M_F=4$ at threshold is
     shown as a function of magnetic field. (f) The scattering length
     (in units of $a_0$) for $^{87}\text{Rb}$  collisions in the lowest
     channel with total $M_F=2$ at threshold is shown as a function of
     magnetic field. (b)-(e) zoom in on the 4 narrow resonance
     features of the $^{87}\text{Rb}$  collisions near $406$G, $686$G, $911$G, and $1007$G, respectively.}
\end{figure*}

 We use the potential energy functions for rubidium developed by Strauss et
 al.~\cite{strauss_hyperfine_2010} without modification. For both isotopes there are
 small, yet significant differences between the scattering lengths reported in
 Ref.~\cite{strauss_hyperfine_2010} and those that we calculate using the same
 potential model. We have not been able to determine the source of this
 discrepancy, but the disagreement motivated us to perform further rigorous tests of our log-derivative propagator.  The results of these tests are carried out in Appendix~\ref{RawitcherTest}.

For $^{85}\text{Rb}$, we consider elastic collisions in the lowest
channel with total $M_F =4$. Field values for the zeroes and poles in
the scattering length are given in Table~\ref{tbl:FieldFeatures}. Our
results for MQDT, EIFT, and EDFT compared to the full coupled channels
calculation (CC) are shown in Fig.~\ref{fig:Rb87} (a) where we plot the
scattering length for fields ranging from $0$G to $1200$G. Coupled channels calculations reveal
two broad resonances at $851.755$G and $1070.9$G which are shown
more clearly in the insets. We find that all methods are able to
replicate the general properties of the scattering length, but MQDT is
superior for predicting the positions of resonances and zero
crossings, matching the CC results almost exactly. EDFT does slightly
worse, coming within a few gauss of the CC results, while EIFT is the
least accurate, routinely overestimating the locations of poles and
zeroes by about $20$G. In 2013, Blackley et al. \cite{blackley_rubidium85} experimentally confirmed $17$ Feshbach resonances in optically trapped
$^{85}\text{Rb}$. For the
ground state channel, they report one $s$-wave Feshbach resonance at
$852.3(3)$G with a width $\Delta > 1$G. Our CC and MQDT calculations fall just outside the uncertainty of this measurement. To the best of our knowledge, no experimental measurements of the high-field resonance near $1071$G has appeared in the literature.

For $^{87}\text{Rb}$, we consider elastic collisions in the lowest
channel with total $M_F =2$. Fig. ~\ref{fig:Rb87} (f) plots the
scattering length for fields between $200$G and $1200$G and
Figs. ~\ref{fig:Rb87} (b)-(e) zoom in on each of the four resonance
features. Similar to the $^{85}\text{Rb}$ case, we find that MQDT
almost exactly reproduces the results of coupled channels calculation while
EIFT overestimates the positions of poles and EDFT comes within $10$G
of the CC results. Turning to experimental data, in 2002, Marte et al. \cite{marte_rubidium87} observed more than 40
resonances in rubidium 87 for magnetic fields between $0.5$G and $1260$G for various spin mixtures in the lower hyperfine ground state to an
accuracy of $30$mG. For the ground state entrance channel, they
report $s$-wave Feshbach resonances at $406.2(3)$G, $685.4(3)$G, $911.7(4)$G, and $1007.3(4)$G. A more recent study conducted by Ref.~\cite{volz_rubidium87} places the high field resonance at
$1007.40(4)$G and measures a zero crossing in the scattering length at $1007.60(3)$G. We find that the values predicted by our CC and MQDT calculations are nearly within the experimental uncertainty of both Refs.~\cite{marte_rubidium87} and ~\cite{volz_rubidium87}.

\subsection{Cesium}
\begin{figure}[t]
   \centering
   \includegraphics[width=\linewidth]{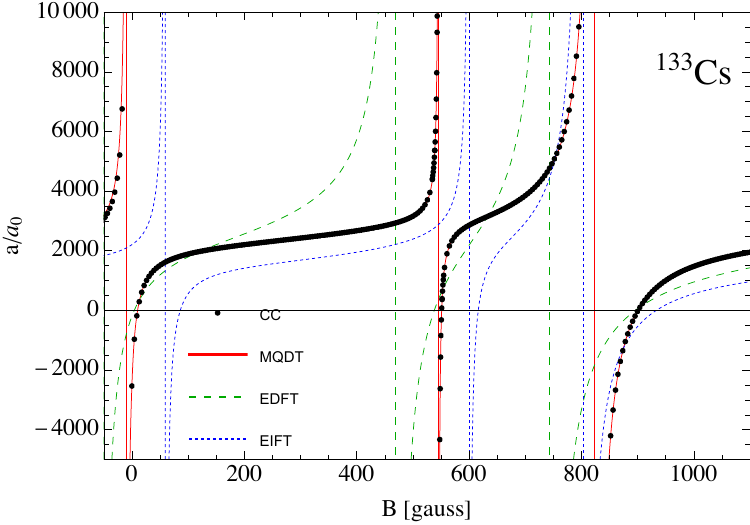}
   \caption{\label{fig:Cs133} (color online) The scattering length (in units of $a_0$) for $^{133}\text{Cs}$
     collisions in the lowest channel with total $M_F=6$ at threshold is shown as a function of magnetic field.}
\end{figure}

For cesium, we use the MLR potentials of ~\cite{baldwin2012improved},
but with modifications as discussed in Section~\ref{subsec:STPots}. We modify the long-range behavior of the potential to more rapidly converge to the functional form of $V_{\text{LR}}(R)$ in Eq.~\ref{eq:vlr} by using a switching function $f(R)$ that vanishes for for $R\lesssim R_{\text{LR}}-\delta R$ and goes to unity for $R\gtrsim R_{\text{LR}} + \delta R$,
\begin{equation}
V^{(S)}(R)=V_S^{(\text{MLR})}(R)(1-f(R)) + f(R)V_{\text{LR}}(R),
\end{equation}
where the switching function is
\begin{equation}
\label{eq:switching}
    f(R)=\frac{1}{2} \left( \tanh{\left( \frac{R - R_{\text{LR}}}{\delta R} \right)}+1 \right).
\end{equation}
We choose $\delta R = 0.5 a_0$ and $R_{\text{LR}}=38 a_0$ to ensure that when the boundary condition Eq.~(\ref{eq:srbc}) determining ${\mathbf K}_{\text{sr}}$ is applied in at $R_f=40 a_0$, the reference functions $\hat{f}(R)$ and $\hat{g}(R)$ are valid solutions to the Schr\"odinger equation in each channel. Without the switching function, there is little hope of finding agreement between the CC and MQDT calculations for this particular MLR potential. Choosing a smaller $R_{\text{LR}}$ gives better agreement between the CC and MQDT calculations, but also dramatically changes the values of $a_S$, $a_T$, the background scattering length, and the resonance positions---so much so that tuning the parameters $V_{c}^{(S)}$ in order to bring $a_S$ and $a_T$ back in line with accepted values becomes difficult. If better agreement with CC calculations is desired, either a more detailed re-parameterization of the potential model is required, or a different model should be used.

The switching function turns out to be unnecessary for the case of lithium, despite the fact that the lithium MLR potentials exhibit similar slow convergence to the form of $V_{\text{LR}}$ (See Fig.~\ref{fig:pVConv}).  We speculate that this is likely because the lower reduced mass of the lithium dimer leads to a correspondingly slower phase accumulation in the asymptotic region. 

As with lithium, we adjust the short-range behavior of the MLR potentials by
adding a quadratic term given by~\ref{eq:potshift}. We first adjust the parameters 
$V_{c}^{(S)}$ to reproduce the scattering lengths reported in 
Ref.~\cite{chin_precision_2004}, then make further adjustments to best 
reproduce the positions of the three $s$-wave resonances reported in 
Ref.~\cite{berninger_feshbach_2013}. It is not possible to reproduce all three resonance positions by tuning only $V_{c}^{(S)}$, and a full re-parameterization of the potential is beyond the scope of this work. 

Field values for zeroes and pole positions in the scattering length are listed in
Table~\ref{tbl:FieldFeatures}. In Fig.~\ref{fig:Cs133}, we plot the
scattering length, showing 3 $s$-wave resonances for magnetic field
ranging from $-50$G to $1100$G. Comparing MQDT, EDFT, and EIFT to
the converged CC calculations shows that while all three methods are
able to reproduce the qualitative features of $a(B)$, MQDT by far is
the most successful at replicating the locations of resonances and
zero crossings, agreeing to within $1$G. Conversely, EIFT overshoots
the resonances near $-10$G and $545$G by about $50$G, and
underestimates the resonance near $820$G by $20$G. EDFT does
even worse, undershooting the three resonances by about $80$G. However, EDFT does slightly better at predicting the locations of
zero crossing, matching the CC calculations to within $10$G. 

The low-field (i.e., $B\lesssim250$G) resonances of cesium atoms have been studied by several groups ~\cite{vuletic1999observation, chin_2000,chin_precision_2004, gustavsson_cesium, leo_2000,chin_2003,lee_2007,weber_2002,mevznarvsivc2019cesium}. In 1999, Vuletic et al~\cite{vuletic1999observation} observed a low-field resonance 
in the total $M_F=6$ block.  They found a zero and a pole at the following positions: $17.0(2)$G and $30(3)$G. Subsequently, Refs.~\cite{chin_2000,gustavsson_cesium} reported values of $(17.064 \pm 0.056)$G and $17.119(2)$G, respectively, for the position of the zero-crossing in the scattering length. More recently, Ref.~\cite{mevznarvsivc2019cesium} find the zero-crossing to be at $17.26(20)$G. Our CC and MQDT calculations differ from this latest experimental value by $\sim7$G. The discrepancy may be improved by employing interaction potentials such as the M2012 model of Ref.~\cite{berninger_feshbach_2013}.

The zero at $17$G has been used \cite{weber_2002,gustavsson_cesium} to prepare a Bose-Einstein condensate (BEC) of cesium atoms in the ground state. This feature is associated with a broad Feshbach resonance near $-11$G. Physically, a resonance at $-|B|$ corresponds to one at $|B|$ with the spin projections of each atom reversed in sign \cite{chin_feshbach_2010}. In this case, the negative resonance at $-11.7$G in the $M_F=+6$ block corresponds to a positive resonance at $11.7$G in the $M_F=-6$ block, which as been measured by Ref.~\cite{chin_precision_2004}. Other theoretical models predict a location of $-11.1(6)$G~\cite{lange_2009} or $-12$G~\cite{berninger_feshbach_2013} for this low-field $s$-wave resonance. Comparing our results to these values, we find that both the coupled-channels calculation and MQDT overestimate this resonance position by $\sim 3$G. 

Berninger et al.~\cite{berninger_feshbach_2013} have explored the high-field physics  of  
ultracold cesium collisions. Using
trap-loss spectroscopy, they observed two broad loss features around
$549$G and $787$G, which correspond to $s$-wave resonances, and
a zero crossing in the scattering length at $553.73(2)$G. Again, we see a discrepancy between the available experimental data and our calculations. The coupled-channels calculation and MQDT underestimate the first resonance position and the zero-crossing by a few gauss, and overestimate the latter resonance position by almost $50$G. This is a shortcoming of the MLR potential developed in Ref.~\cite{baldwin2012improved} for cesium, and we expect significantly better agreement in future calculations using improved potential models such as the M2012 potential of Ref.~\cite{berninger_feshbach_2013}, which was specifically developed to describe experimental data at \emph{both} low and high fields. 

\section{Concluding Discussion}
\begin{figure}[b]
  \includegraphics[width=3.4in]{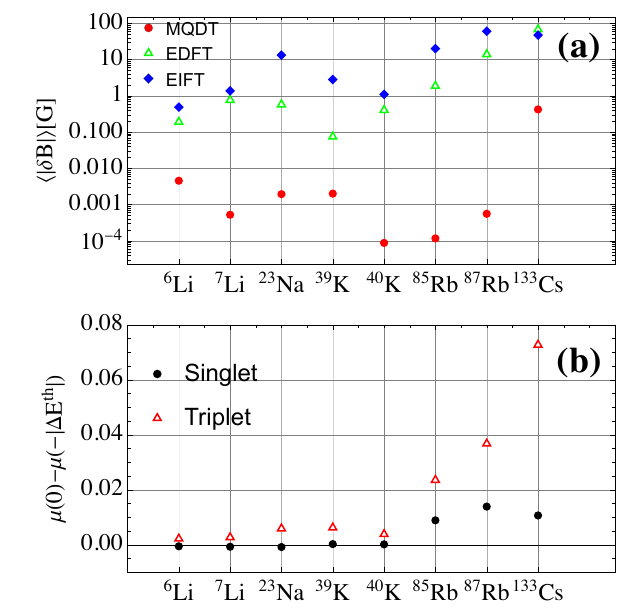}
  \caption{\label{fig:QDsErrors} (color online) Panel (a) shows the average absolute error (in gauss) of resonance positions for each atomic species. Data for MQDT (red circles), EDFT (green triangles) and EIFT (blue diamonds) are all shown on a log scale.  Panel (b) shows the variation of the singlet and triplet quantum defects over the necessary range of energy required for the energy-dependent frame transformation calculation.}
  \end{figure}

The accuracy of the EIFT, EDFT and MQDT calculations depend on a number of factors that we will now attempt to untangle. In Panel (a) of Fig.~\ref{fig:QDsErrors}, we show the mean absolute error (in gauss) of magnetic Feshbach resonance positions for each atomic species. The error is defined for each of the resonance (pole) positions in Table~\ref{tbl:FieldFeatures} simply as
\begin{equation}
\delta B = \lpipe B_{\text{pole}}^{\text{type}} - B_{\text{pole}}^{\text{CC}} \rpipe,
\end{equation}
where ``type" stands for any of the MQDT, EDFT or EIFT calculations, and CC stands for the coupled channels calculation, which we have ensured are fully converged. We have taken care to compute a higher density of points in the vicinity of resonance poles and zeros of the scattering length. An interpolating function is used to identify the zeroes of $1/a(B)$ as pole locations, accelerating the convergence of the CC calculations in particular when searching for these features.

Let us first consider the elements of the MQDT calculation that may limit its accuracy. First, and likely the most significant contributor to error, is the fact that the MLR potentials themselves converge rather slowly to their asymptotic form, as illustrated in Fig.~\ref{fig:pVConv}.  Therefore, the reference functions $\hat{f}(R)$ and $\hat{g}(R)$, which are solutions to the Schr\"odinger equation in a potential $V_{\text{LR}}$, are not perfect solutions to the Schr\"odinger equation in $V^{\text{MLR}}$. Even slight differences in the long-range potentials can lead to a substantial difference in the resonance position.  Secondly, the reference functions themselves are computed numerically and any error in their computation is inherited by ${\bf K}_{\text{sr}}$. The MQDT calculations (red circles) are typically several orders of magnitude more accurate than either of the frame transformation calculations, but MQDT performs most poorly for $^6\text{Li}$ and $^{133}\text{Cs}$. We believe that this is primarily caused by the slow convergence of the MLR potentials to the asymptotic form $V_{\text{LR}}$ of Eq.~\ref{eq:vlr}, as shown in Fig.~\ref{fig:pVConv}. Without the switching function Eq.~\ref{eq:switching}, the MQDT calculation for $^{133}\text{Cs}$ is significantly poorer. Likewise, without extending the matching radius $R_m$ out to about $55a_0$ for lithium, as discussed in subsection~\ref{subsec:STPots}, the performance of MQDT is significantly worse than what is shown. For further improvements, we recommend using a different potential energy model with faster convergence to $V_{\text{LR}}$.

The frame transformation calculations rely upon the singlet and triplet quantum defects $\mu_S$ which are plotted in Fig.~\ref{fig:QD}. Panel (b) of Fig.~\ref{fig:QDsErrors} shows the overall variation of the quantum defects over the total energy range required for the EDFT calculation, as prescribed by Eq.~(\ref{eq:averageE}). The first feature to note is that as a rule, the energy dependence of the triplet quantum defects is greater than that of the singlet defects.  This is sensible since the separation of energy and length scales is more robust for the comparatively deep singlet channel, leading to weaker energy dependence in $\mu_0(E)$ compared to $\mu_1(E)$. The second feature to note is that the performance of the frame transformation calculations is strongly correlated to the energy dependence of the quantum defects themselves. In general, the heavier the species, the greater the sensitivity to energy displayed by the quantum defects. This is because the hyperfine-Zeeman splitting increases with atomic mass. The one exception to this trend is $^{40}\text{K}$, in which there are only three collision thresholds with total $M_F=-7$, and the range of energies over which one must evaluate the quantum defects is considerably smaller. 

Only the MQDT calculation is able to reliably reproduce the position of every resonance pole to less than the width of the resonance. See, for example, Panels (a)-(d) of Fig.~\ref{fig:Rb87} showing the individual $s$-wave resonances in $^{87}\text{Rb}$. The EDFT provides a significant improvement over EIFT in all cases, except for cesium. For example, see the wide resonance in $^{23}\text{Na}$ near $915$G shown in Fig.~\ref{fig:Na23}, the resonances near $26$G and $752$G in $^{39}\text{K}$ shown in Fig.~\ref{fig:K40}, or even the two resonances shown in the insets of Fig.~\ref{fig:Rb87}(a) for $^{85}\text{Rb}$.

To conclude, we have conducted a comprehensive study of ultracold homonuclear collisions for eight alkali species, applying three variations of multichannel quantum defect theory that differ in how they characterize the short-ranged K-matrix, ${\bf K}_{\text{sr}}$. We have attempted to untangle various sources of error, both among the calculations themselves, and with experiment. We have quantitatively demonstrated how the frame transformation calculations become rather unreliable for the heavier species with large hyperfine-Zeeman splittings, while MQDT remains robust provided that the singlet and triplet potentials converge sufficiently quickly to the long-range form of $V_{\text{LR}}$. We hope to perform calculations in the future that extend this work to higher partial waves and include the weak magnetic dipole-dipole coupling.  A still more comprehensive study of inelastic processes is also within reach.
\begin{acknowledgments}
We thank Chris H. Greene for guidance in the early stages of this work. 
\end{acknowledgments}
\appendix
\section{Numerical testing of the log-derivative propagator}
\label{RawitcherTest}
Numerical discrepancies between (some of) our calculated singlet and triplet
scattering lengths and those reported in the literature, particularly
for rubidium, spurred us to
conduct further testing of our computer code.  Here, we present
calculations using the two-channel test model of
Ref.~\cite{rawitscher1999comparison}. 

\begin{figure}[t]
   \centering
   \includegraphics[width=8cm]{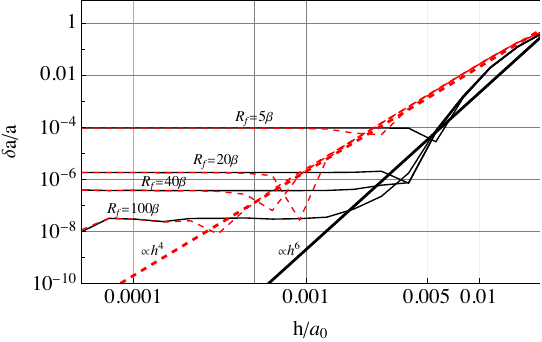}
   \caption{\label{fig:rawitscher} (color online) The scattering length (in units of $a_0$) for $^6$Li
     collisions in the lowest channel with total $M_F=0$ at threshold is shown as a function of magnetic field.}
\end{figure}

The authors of
Ref.~\cite{rawitscher1999comparison} compare three robust methods
commonly used for solving coupled channels problems: (1) the integral
equation method (IEM)~\cite{gonzales1997integral}, (2) the finite element~\cite{burke1996multichannel,bathe1976numerical}
eigenchannel R-matrix~\cite{greene1983atomic} propagator, and (3) the
Gordon algorithm~\cite{gordon1969new}. Of these three methods, the greatest stability is
achieved by the IEM, which---when used with a perturbative long-range
correction---gives the scattering length to 11 significant figures
$a(\infty) = 851.98171574$. We directly compare our calculation of the
scattering length, $a$, to $a(\infty)$. and plot $(\delta a) / a = (a-a(\infty))/a(\infty)$ as a function of
step size $h/a_0$. Our implementation of Johnson's
log-derivative propagator~\cite{johnson1973multichannel} uses Richardson extrapolation with step
doubling~\cite{press_etal_nr}, which greatly improves the convergence
scaling with step size from $h^4$ to $h^6$. We therefore present two sets of calculations
in Fig.~\ref{fig:rawitscher}.  One with Richardson extrapolation (solid
black curves), and one without (red dashed curves).  The thick red dashed line
indicates $h^4$ scaling, while the thick black solid line indicates
$h^6$ scaling.  Calculations for various values of $R_f$
(where the matching to Bessel functions is made) are shown.
These results clearly demonstrate the improved scaling, the dependence
on $R_f$, and the dependence on step size $h/a_0$.  We
typically use $N=10^7$ integration steps and integrate out to
$R_f = 20 \beta \lesssim 4000 a_0 $ in all our log-derivative
calculations. Calculations including higher partial waves will require a larger matching radius. Based on the results shown in Fig.~\ref{fig:rawitscher},
we expect about 6 significant figures in the scattering length.

\section{Numerical Solution for the Reference Functions $\hat{f}$ and
  $\hat{g}$.}
\label{numericalfg}
The reference functions $\hat{f}_i(R)$ and $\hat{g}_i(R)$ (See Eq. (\ref{eq:vdw-ref-funs})) are obtained from the Milne equation,
Eq.~(\ref{eq:alpha-fun}), with WKB-like boundary conditions Eq.~(\ref{eq:alpha-bc1}) imposed at $R_x\approx 0.07 \beta$,
deep in the long-range reference potential $V_{\text{LR}}(R)$.  MQDT requires that we calculate the \emph{single channel} parameter
$\cot{(\gamma)}$ via Eq.~(\ref{eq:gamma}) as a function of energy, anticipating that for elastic
collisions in the lowest open channel, both the MQDT and EDFT calculations require us to evaluate $\cot{(\gamma)}$ at
negative energies equal to the separation of two-atom
hyperfine-Zeeman splitting at magnetic field strengths or order $1000$G. It is therefore necessary to calculate the
reference functions out to an asymptotic matching distance $R_f$ that lies well into the classically forbidden region where the
solution $\alpha(R)$ grows without bound.  In order to avoid numerical overflow, we perform a simple variable
transformation of the Milne equation by letting $\alpha(R) = e^{x(R)}$, leading to the following nonlinear equation for $x(R)$ 
\begin{equation}
  \label{eq:modified-milne}
x'' + (x')^2 = e^{-4x}-k^2.
\end{equation}
We solve Eq.~\ref{eq:modified-milne} by fourth order Runge-Kutta (RK4) with repeated step doubling and Richardson
extrapolation (repeated twice)~\cite{zlatev_explicit_2020}.  To be clear, we define $y_1(R) = x$, and $y_2(R)=x'(R)$.
Then apply the usual RK4~\cite{press_etal_nr} procedure to the set of coupled equations:
\begin{equation}
  \begin{pmatrix}
    y_1'\\
    y_2'
  \end{pmatrix}
  =
  \begin{pmatrix}
    y_2\\
    e^{-4y_1} - k^2 -y_2^2
  \end{pmatrix}
\end{equation}
The log-derivatives of the reference functions are easily computed:
\begin{align}
  \frac{\hat{f}'}{\hat{f}} &= y_2 + e^{-2 y_1} \cot{(\phi(R) + \phi_L)} \\
  \frac{\hat{g}'}{\hat{g}} &= y_2 - e^{-2 y_1} \tan{(\phi(R) + \phi_L)},
\end{align}
where $\alpha(R)=e^{y_1(R)}$. This provides a numerically stable way
to compute all necessary quantities, including the phase
standardization from Eq.~(\ref{eq:tanphi}).

\bibliography{MQDTRefs}

\end{document}